\documentstyle[12pt]{article}

\begin{document}
\input{psfig.sty}
\begin{flushright}
\baselineskip=12pt
hep-ph/0210436 \\
ANL-HEP-PR-02-092\\
\end{flushright}

\begin{center}
\vglue 1.2cm
{\Large\bf  Low Energy 6-Dimensional N=2 Supersymmertric  SU(6) Models on $T^2$ Orbifolds}
\vglue 1.2cm
{\Large   Jing Jiang$^a$,
 Tianjun Li$^b$ and Wei Liao$^c$ }
\vglue 0.6cm
{$^a$HEP Division, Argonne National Laboratory, \\
9700 S. Cass Avenue, Argonne, IL 60439\\
$^b$School of Natural Sciences, Institute for Advanced Study, \\
Einstein Drive, Princeton, NJ 08540\\
$^c$The Abdus Salam International Center for Theoretical Physics,\\
Strada Costiera 11, 34014 Trieste, Italy}
\end{center}

\vglue 1.2cm
\begin{abstract}
We propose low energy 6-dimensional $N=2$ supersymmetric $SU(6)$ models
on $M^4\times T^2/(Z_2)^3$ and $M^4\times T^2/(Z_2)^4$, where
the orbifold $SU(3)_C\times SU(3)$ model can be embedded on the boundary
4-brane.
For the zero modes, the 6-dimensional $N=2$ supersymmetry and the $SU(6)$ gauge
symmetry are broken down to the 4-dimensional $N=1$ supersymmetry and
the $SU(3)_C\times SU(2)_L\times U(1)_Y\times U(1)'$ gauge symmetry by orbifold
projections. 
In order to cancel the anomalies involving
at least one $U(1)'$, we add extra exotic particles. We also study the
anomaly free conditions and present some anomaly free models.
The gauge coupling unification can be achieved at
$100\sim 200$ TeV if the compactification
scale for the fifth dimension is $3\sim 4$ TeV. The proton decay
problem can be avoided by putting the quarks and leptons/neutrinos on
different 3-branes. 
And we discuss how to break the
$SU(3)_C\times SU(2)_L\times U(1)_Y\times U(1)'$ gauge symmetry,
solve the $\mu$ problem, and generate the
$Z-Z'$ mass hierarchy naturally by using the geometry. 
The masses
 of exotic particles can be at the order of 1 TeV after
the gauge symmetry breaking. 
We also forbid the dimension-5 operators for the neutrino masses
 by $U(1)'$ gauge symmetry,
 and the realistic left-handed neutrino masses 
can be obtained via non-renormalizable
terms.

\end{abstract}

\vspace{0.5cm}
\begin{flushleft}
\baselineskip=12pt
October 2002\\
\end{flushleft}
\newpage
\setcounter{page}{1}
\pagestyle{plain}
\baselineskip=14pt

\section{Introduction}

Grand Unified Theory (GUT) has a two-fold meaning:
(1) Gauge coupling unification, {\it i.e.},
the gauge couplings for the Standard Model (SM)  gauge groups
$SU(3)_C\times SU(2)_L\times U(1)_Y$ are equal at GUT scale;
(2) Fermion unification, {\it i.e.}, the SM fermions of each generation elegantly
fit into the $\bar{5}+10$ representation of $SU(5)$ group, or the spinor $16$ representation of
 $SO(10)$ group if we included the right handed neutrinos. As we know,
GUT gives us an simple
 and elegant understanding of the quantum numbers of quarks and leptons,
and the success of gauge coupling unification in the Minimal Supersymmetric
Standard Model (MSSM) strongly supports
 this idea.   The radiative electroweak symmetry breaking can also 
be realized in the supersymmetric GUT due to the large top quark Yukawa coupling.
Another impressive
success of the supersymmetric GUT is that including the radiative corrections,
GUT gives  
the correct weak mixing angle, as observed in the electroweak (EW) scale
experiments. Therefore,
 the Grand Unified Theory at high energy scale has
been widely accepted. However, there are some problems in GUT: 
 the grand unified
gauge symmetry breaking mechanism, the doublet-triplet splitting problem, 
the proton decay problem, and the fermion mass relation
 problem ($m_e/m_{\mu} = m_d/m_s$), etc.

About two years ago, a new scenario was proposed to address above questions in 
GUT~\cite{kawa, alot, LTJ1, LTJ2}.
 The key point is that the supersymmetric GUT
model exists in 5 or higher dimensions and is broken down to the
4-dimensional 
$N=1$ supersymmetric Standard-Model-like Model for
 the zero modes due to the 
discrete symmetries in the neighborhoods of the branes
or on the extra space manifolds, which
become non-trivial constraints on the multiplets and gauge generators in 
GUT~\cite{kawa, alot, LTJ1, LTJ2}. 
In addition, the discrete symmetry may not act freely on
the extra space manifold. When the discrete symmetry 
does not act freely on the extra space manifold, there exists a brane at 
each fixed point, line or hypersurface, where only part of 
the gauge symmetry and supersymmetry might be preserved and
the SM fermions can be located~\cite{kawa, alot, LTJ1, LTJ2}.
The attractive models have been constructed explicitly, in which
the supersymmetric 5-dimensional and 6-dimensional 
GUT models are broken down to
the 4-dimensional $N=1$
 supersymmetric $SU(3)\times SU(2) \times U(1)^{n-3}$
models, where $n$ is the rank of the GUT groups, through the 
compactification on various orbifolds and manifolds. 
The GUT gauge symmetry breaking can be achieved, and the doublet-triplet
splitting problem, the proton decay problem and
the fermion mass relation
 problem ($m_e/m_{\mu} = m_d/m_s$) can be solved neatly by the
discrete symmetry projections, although the fermion unification
might be given up.

In addition to the orbifold GUT models,  the low energy partial unification
$SU(3)_C\times SU(3)$ model on the space-time $M^4\times S^1/(Z_2\times Z_2')$
has been proposed recently,
where quarks are put on the 3-brane at the fixed point on which only
the Standard Model gauge symmetry is preserved~\cite{LL1, hn, kdw}.  In this model,
the desirable weak mixing angle $\sin^2\theta_W=0.2312$
at $m_Z$ scale can be generated naturally and the $U(1)_Y$ charge quantization
can be obtained due to the gauge invariant of the
Yukawa couplings and anomaly free conditions.
We also showed that from the point of view of the gauge
coupling runnings, a complete gauge coupling
unification might occur at $10^{6}$ GeV if the
compactification scale of the fifth dimension ($1/R$) is
$10^4$ GeV for the supersymmetric model
without embedding it into a GUT model~\cite{LL1}. 
However, there exists the $\mu$ problem.
And the left-handed neutrinos can obtain masses via
the dimension-5 operators $y_{\nu ij} L_i L_j H_u H_u/M_*$,
where $M_*$ is the fundamental scale (or cut-off scale) in
the model which is about $10^{6}$ GeV.
Therefore, the masses for the left-handed neutrinos
can not be realistic unless  $y_{\nu ij}$ is very small,
at the order of $10^{-7}$ or less.

On the other hand, the possibility of an extra $U(1)'$ gauge symmetry 
is well-motivated in superstring constructions~\cite{string},
 grand unified theories~\cite{review},
and in  models of dynamical symmetry breaking~\cite{DSB}.
In supersymmetric models, an extra $U(1)'$ can provide an elegant
solution to the $\mu$
problem~\cite{muprob1,muprob2}, with an effective $\mu$ parameter
generated by the vacuum expectation value (VEV) of the Standard Model singlet
field $S$ which breaks the $U(1)'$ gauge symmetry. This is somewhat similar to
the effective $\mu$ parameter in the Next to Minimal Supersymmetric Standard
Model (NMSSM)~\cite{NMSSM}. However, with a $U(1)'$ the extra discrete
symmetries and their associated cosmological problems typically associated
with the NMSSM are absent. In superstring-motivated models it is often
the case that the electroweak and $U(1)'$ breaking are both driven by the
soft supersymmetry breaking parameters, so one typically expects the mass
of the $U(1)'$ gauge boson $Z'$
to be of the same order as the electroweak scale~\cite{string}.
However, there are stringent limits on $Z'$ mass from direct searches
during Run I at the Tevatron~\cite{explim} and from indirect precision tests at the
$Z$-pole, at LEP 2, and from weak neutral current experiments~\cite{indirect}.
The constraints depend on the particular $U(1)'$ couplings, but in typical
models one requires that $M_{Z^{\prime}} > (500-800) $ GeV
and the $Z-Z'$ mixing angle
$\theta_{Z-Z^{\prime}}$ to be smaller than a few $\times 10^{-3}$.
Thus, how to explain the $Z-Z^{\prime}$ mass hierarchy is an interesting question.
Recently,  the supersymmetric model with a secluded $U(1)^{\prime}$-breaking
sector was proposed, where the squark and slepton spectra can mimic those of the MSSM,
the electroweak breaking is actually driven by the relatively large $A$
terms, and a large $Z'$ mass can be generated by the VEVs of additional
SM singlet fields that are charged under the $U(1)'$~\cite{JPT}.
In this scenario, the superpotential for the Higgs is
\begin{eqnarray}
W &=& h S H_d H_u + \lambda S_1 S_2 S_3 ~,~\,
\end{eqnarray} 
where $S$ and $S_i$ are the Standard Model singlets, $h$ and $\lambda$
are coupling constants. In order to generate the $Z-Z'$ mass hierarchy,
one has to choose $h\sim 10 \lambda$, which becomes the Yukawa couplings
hierarchy.

In this article, 
we propose the low energy 6-dimensional $N=2$ supersymmetric $SU(6)$ model
on the space-time $M^4\times T^2/(Z_2)^3$ and $M^4\times T^2/(Z_2)^4$, where
the orbifold $SU(3)_C\times SU(3)$ model can be embedded on the 
4-brane at $z=0$, and the desirable weak mixing angle $\sin^2\theta_W=0.2312$
at $m_Z$ scale can be generated naturally and the $U(1)_Y$ charge quantization
for the Standard Model fermions 
can be obtained due to the gauge invariant of the Yukawa couplings and anomaly free conditions.
For the  zero modes, the 6-dimensional $N=2$ supersymmetry and the $SU(6)$ gauge
symmetry are broken down to the 4-dimensional $N=1$ supersymmetry and
the $SU(3)_C\times SU(2)_L\times U(1)_Y\times U(1)'$ gauge symmetry by orbifold
projections. We
present the parity assignment and masses for the bulk gauge fields and
Higgs fields on the 4-brane at $z=0$, and the number of the 4-dimensional
supersymmetry and gauge symmetry on the 3-branes at the fixed points and
on the 4-branes on the fixed lines.
 In order to cancel the anomaly involving
at least one $U(1)'$, we add extra exotic particles which
are vector-like under the Standard Model gauge symmetry.
We examine the anomaly free conditions and give
some anomaly free models.
In addition, the gauge coupling unification can be achieved at
$100\sim 200$ TeV if the compactification
scale for the fifth dimension ($1/R_1$) is $3\sim 4$ TeV. The proton decay
problem can be avoided by putting the quarks and leptons/neutrinos on
different 3-branes. Moreover, we discuss the 
$SU(3)_C\times SU(2)_L\times U(1)_Y\times U(1)'$ gauge symmetry breaking,
solve the $\mu$ problem and generate the
$Z-Z'$ mass hierarchy naturally because
we can have $h \sim 10 \lambda$ if we
put the Higgs $H_u$, $H_d$ and $S_i$ on the 4-brane at $z=0$
and put the $S$ on the 3-brane at the fixed point.
 The masses for the extra exotic particles
 can be at the order of 1 TeV after the gauge symmetry breaking.
In particular, we consider the neutrino masses in detail. We forbid the dimension-5
operators by $U(1)'$ gauge symmetry which might give very large
masses to the left handed neutrinos unless the couplings are at the order
of $10^{-8}$ or less, and the
correct active neutrino masses can be obtained via the non-renormalizable
terms naturally.

This paper is organized as follows. In Section 2, we review the 6-dimensional $N=2$ supersymmetric
gauge theory and explain our convention. 
In Section 3,
 we consider the low energy 6-dimensional $N=2$ supersymmetric $SU(6)$ gauge 
unification theory on the space-time $M^4\times T^2/(Z_2)^3$.
And in Section 4, we study the low energy 6-dimensional $N=2$ supersymmetric $SU(6)$ gauge 
unification theory on the space-time $M^4\times T^2/(Z_2)^4$.
The discussion and conclusion are given in Section 5.

\section{$N=2$ 6-Dimensional Supersymmetric Gauge Theory and Convention}

$N=2$ supersymmetric gauge theory in 6-dimension has 16 real supercharges,
corresponding to $N=4$ supersymmetry in 4-dimension. So, only the
vector multiplet can be introduced in the bulk. 
In terms of 4-dimensional
$N=1$ supersymmetry language, it contains a vector multiplet $V(A_{\mu}, \lambda_1)$,
and three chiral multiplets $\Sigma_5$, $\Sigma_6$, and $\Phi$. All 
of them are in the adjoint representation of the gauge group. In addition,
the $\Sigma_5$ and $\Sigma_6$ chiral multiplets
contain the gauge fields $A_5$ and $A_6$ in
their lowest components, respectively.

In the Wess-Zumino gauge and 4-dimensional $N=1$ language, the bulk action 
is~\cite{NAHGW}
\begin{eqnarray}
  S &=& \int d^6 x \Biggl\{
  {\rm Tr} \Biggl[ \int d^2\theta \left( \frac{1}{4 k g^2} 
  {\cal W}^\alpha {\cal W}_\alpha + \frac{1}{k g^2} 
  \left( \Phi \partial_5 \Sigma_6 - \Phi \partial_6 \Sigma_5
  - \frac{1}{\sqrt{2}} \Phi 
  [\Sigma_5, \Sigma_6] \right) \right) 
\nonumber\\
&& + {\rm H.C.} \Biggr] 
  + \int d^4\theta \frac{1}{k g^2} {\rm Tr} \Biggl[ 
  \sum_{i=5}^6 \left((\sqrt{2} \partial_i + \Sigma_i^\dagger) e^{-V} 
  (-\sqrt{2} \partial_i + \Sigma_i) e^{V} + 
   \partial_i e^{-V} \partial_i e^{V}\right)
\nonumber\\
  && \qquad \qquad \qquad
  + \Phi^\dagger e^{-V} \Phi e^{V}  \Biggr] \Biggr\} ~.~\,
\label{eq:5daction}
\end{eqnarray}
And the gauge transformation is given by
\begin{eqnarray}
  e^V &\rightarrow& e^\Lambda 
    e^V e^{\Lambda^\dagger}, \\
  \Sigma_i &\rightarrow& e^\Lambda (\Sigma_i - \sqrt{2} \partial_i) 
    e^{-\Lambda}, \\
  \Phi &\rightarrow& e^\Lambda \Phi e^{-\Lambda}~,~\,
\end{eqnarray}
where $i=5, 6$.

Our convention is the same as that of Ref.~\cite{LTJ1}. We consider 
the 6-dimensional space-time which can be factorized into a product of the 
ordinary 4-dimensional Minkowski space-time $M^4$, and the torus $T^2$
which is homeomorphic to $S^1\times S^1$. The corresponding
coordinates for the space-time are $x^{\mu}$, ($\mu = 0, 1, 2, 3$),
$y\equiv x^5$ and $z\equiv x^6$. 
The radii for the circles along the $y$ direction and $z$ direction are
$R_1$ and $R_2$, respectively.
We also define $y'$ and $z'$ by $y' \equiv y-\pi R_1/2$ 
and $z' \equiv z- \pi R_2/2$. In addition,
 we assume that the gauge fields are in
the bulk, and the SM fermions and Higgs particles are on the 
4-brane on the 
 fixed line (boundary) or on the 3-brane at the fixed point
in the extra space orbifold.

In this paper, 
the orbifold $T^2/(Z_2)^3$ are defined by $T^2$ moduloing three equivalent classes
\begin{equation}
y\sim -y~,~ z\sim -z~,~z'\sim -z'~.~\,
\end{equation} 
Precisely speaking, our orbifold $T^2/(Z_2)^3$ is $S^1/Z_2 \times S^1/(Z_2\times Z_2')$.
The fixed points are $(y=0, z=0)$, $(y=0, z=\pi R_2/2)$,
$(y=\pi R_1, z=0)$ and $(y=\pi R_1, z=\pi R_2/2)$,
and the fixed lines are $y=0$, $z=0$, $y=\pi R_1$ and
$z=\pi R_2/2$. And the extra space orbifold is a rectangle.

In addition, the orbifold $T^2/(Z_2)^4$ are obtained by $T^2$ moduloing the equivalent classes
\begin{equation}
 y\sim -y~,~ z\sim -z~,~ y'\sim -y'~,~ z'\sim -z'~.~\,
\end{equation}
The four fixed points are
$(y=0, z=0),$ $(y=0, z=\pi R_2/2),$ $(y=\pi R_1/2, z=0)$, and 
$(y=\pi R_1/2, z=\pi R_2/2)$, and the fixed lines
are $y=0$, $z=0$, $y=\pi R_1/2$ and $z=\pi R_2/2$.

For a generic bulk field $\phi(x^{\mu}, y, z)$,
we can define four parity operators $P^y$, $P^z$, 
$P^{y'}$, and $P^{z'}$~\footnote{For the model on $T^2/(Z_2)^3$ orbifold, we just neglect
the $P^{y'}$ parity operator.}
\begin{eqnarray}
\phi(x^{\mu},y, z)&\to \phi(x^{\mu},-y, z )=P^y \phi(x^{\mu},y, z)
 ~,~\,
\end{eqnarray}
\begin{eqnarray}
\phi(x^{\mu},y, z)&\to \phi(x^{\mu}, y, -z )=P^z \phi(x^{\mu},y, z)
 ~,~\,
\end{eqnarray}
\begin{eqnarray}
\phi(x^{\mu},y', z')&\to \phi(x^{\mu},-y', z' )=P^{y'} \phi(x^{\mu},y', z')
 ~,~\,
\end{eqnarray}
\begin{eqnarray}
\phi(x^{\mu},y', z')&\to \phi(x^{\mu}, y', -z' )=P^{z'} \phi(x^{\mu},y', z')
 ~.~\,
\end{eqnarray}

Furthermore, suppose $G$ is a Lie group and $H$ is a subgoup of $G$,
we denote the commutant of $H$ in $G$ as $G/H$, {\it i.e.},
\begin{equation}
G/H\equiv \{g \in G|gh=hg, ~{\rm for ~any} ~ h \in H\}~.~\,
\end{equation}
And if $H_1$ and $H_2$ are the subgroups of $G$,
we define
\begin{equation}
G/\{H_1 \cup H_2\}\equiv \{G/H_1\} \cap \{G/H_2\}~.~\,
\end{equation}
For simplicity, we define
\begin{eqnarray}
G/{P^{y}}~\equiv~ G/\{e, P^{y}\} ~,~\,
\end{eqnarray}
\begin{eqnarray}
G/\{P^{y} \cup P^{z}\}~\equiv~ G/\{e, P^{y}\} \cap G/\{e, P^{z}\}
 ~,~\,
\end{eqnarray}
similarly for the others.

\section{$SU(6)$ Model on $M^4 \times T^2/(Z_2)^3$}

In this section, we discuss the $N=2$ supersymmetric
$SU(6)$ model on the space-time $M^4 \times T^2/(Z_2)^3$.
In particular, on the 4-brane at $z=0$, because of the orbifold projections,
there exist only the
$SU(3)_C\times SU(3) \times U(1)'$ gauge symmetry
and 4-dimensional $N=2$ supersymmetry, where the 
previous orbifold $SU(3)_C\times SU(3)$ model in Refs.~\cite{LL1, hn, kdw} can be embedded naturally.

\subsection{Orbifold $SU(6)$ Breaking, and Particle Spectrum for Gauge and Higgs Fields}

Because in our model, there exist only the 
$SU(3)_C\times SU(3) \times U(1)'$ gauge symmetry
and 4-dimensional $N=2$ supersymmetry on the 4-brane at $z=0$,
we put one pair of Higgs triplets $\Psi_u$ and $\Psi_d$ which transform as
$(1, 3, c_2)$ and $(1, \bar 3, c_1)$ under the
 $SU(3)_C\times SU(3) \times U(1)'$ gauge symmetry on the 4-brane at $z=0$.
We require that $c_1\not=0$, $c_2\not=0$ and $c_1+c_2\not=0$\footnote{If $c_1+c_2=0$,
we can not forbid the brane localized superpotential $\mu H_d H_u$. So, we can not
solve the $\mu$ problem.}. 
We also add three $SU(3)_C\times SU(3)$ singlets,  $\Psi_{S_1}$, 
$\Psi_{S_2}$ and $\Psi_{S_3}$
with $U(1)'$ charges  $-s$, $-s$ and $2s$ respectively
on the 4-brane at $z=0$, where $s\equiv-c_1-c_2$. 
In terms of the 4-dimensional $N=1$ supersymmetry language, the hypermultiplets
$\Psi_u$ and $\Psi_d$ can be decomposed into two pairs of chiral multiplets
$(\Phi_u, \Phi_u^c)$ and $(\Phi_d, \Phi_d^c)$, and
the hypermultiplet $\Psi_{S_i}$ can be decomposed into one pair
of chiral multiplets $(S_i, S^c_i)$ in which $i=1, 2, 3$. Here, the superscript
$c$ means the charge conjugation.
To be explicit, we define
\begin{eqnarray}
\Phi_u \equiv \pmatrix{H_u \cr S_u} ~,~
\Phi_u^c \equiv \pmatrix{H_u^c \cr S_u^c}~,~\,
\end{eqnarray}
\begin{eqnarray}
\Phi_d \equiv \pmatrix{H_d \cr S_d} ~,~
\Phi_d^c \equiv \pmatrix{H_d^c \cr S_d^c}~.~\,
\end{eqnarray}

Moreover, we introduce extra exotic particles to cancel the anomalies, and  
put the Standard Model fermions and extra exotic particles on
the 3-branes at the orbifold fixed points, which will be explained in detail later.

We choose the following matrix representations for the parity operators
$P^y$, $P^{z}$ and $P^{z'}$, 
which are expressed in the adjoint representation of SU(6)
\begin{equation}
P^y={\rm diag}(+1, +1, +1, -1, -1, +1)
\end{equation}
\begin{equation}
P^z={\rm diag}(+1, +1, +1, -1, -1, -1)
~,~ P^{z'}={\rm diag}(+1, +1, +1, +1, +1, +1)~.~\,
\end{equation}

Thus, under the $P^{y}$ and $P^{z}$ parities,
the $SU(6)$ gauge generators $T^A$, where A=1, 2, ..., 35 for $SU(6)$,
are separated into four sets: $T^{a, b}$ are the gauge generators for
 the $SU(3)_C\times SU(2)_L \times U(1)_Y \times U(1)'$ gauge symmetry, $T^{ a, \hat b}$,
$T^{\hat a, b}$, and $T^{\hat a, \hat b}$
 are the other broken gauge generators that
belong to $\{G/P^{y} \cap \{{\rm coset}~ G/P^{z}\}\}$,
$\{\{{\rm coset}~ G/P^{y}\} \cap  G/P^{z}\}$, 
and $\{\{{\rm coset}~ G/P^{y}\} \cap \{{\rm coset}~ G/P^{z}\}\}$,
respectively.
Therefore,
under $P^{y}$, $P^{z}$ and $P^{z'}$, the gauge generators transform as
\begin{equation}
P^{y}~T^{a, B}~(P^{y})^{-1}= T^{a, B} ~,~ 
P^{y}~T^{\hat a, B}~(P^{y})^{-1}= - T^{\hat a, B}
~,~\,
\end{equation}
\begin{equation}
P^{z}~T^{A, b}~(P^{z})^{-1}= T^{A, b} ~,~ P^{z}~T^{A, \hat b}~(P^{z})^{-1}= - T^{A, \hat b}
~,~\,
\end{equation}
\begin{equation}
P^{z'}~T^{A, B}~(P^{z'})^{-1}= T^{A, B} 
~.~\,
\end{equation}

The generators for $SU(6)$ are the $6\times 6$ traceless Hermitian matrix. 
According to the $6\times 6$ matrix, the decompositions of the $SU(6)$
generators $T^{A, B}$ into
the $T^{a, b}$, $T^{ a, \hat b}$, $T^{\hat a, b}$, and $T^{\hat a, \hat b}$ 
are
\begin{eqnarray}
T^{A, B}  = \left(\matrix{ (T^{a, b})_{3\times 3} 
& (T^{\hat{a}, \hat{b}})_{3\times 2}   & (T^{a, \hat{b}})_{3\times 1}  \cr
(T^{\hat{a}, \hat{b}})_{2\times 3} & (T^{a, b})_{2\times 2}
& (T^{\hat{a}, b})_{2\times 1} \cr
(T^{a, \hat{b}})_{1\times 3} & (T^{\hat{a}, b})_{1\times 2}
& (T^{a, b})_{1\times 1}\cr}\right)
~,~ \,
\end{eqnarray}
where $(T^{A, B})_{i\times j}$ means a $i\times j$ matrix.

It is easy to find the  corresponding
generators in $SU(6)$ for 
the generators of the $SU(3)_C$ and $SU(2)_L$ gauge symmetry.
Here,  we write down the explicit
generators for $U(1)_Y$ and $U(1)'$
\begin{eqnarray}
T_{U(1)_Y} &=&  {\rm diag}(0, 0, 0, {1\over 2}, {1\over 2}, -1)~,~\,
\end{eqnarray}
\begin{eqnarray}
T_{U(1)'} &=& {1\over {2 \sqrt 3}} {\rm diag}(1, 1, 1, -1, -1, -1)~.~\,
\end{eqnarray}

Once again, we would like to emphasize that only
 the $SU(6)/P^z=SU(3)_C\times SU(3) \times U(1)'$ gauge
symmetry and 4-dimensional $N=2$ supersymmetry are preserved on the 4-brane at $z=0$.
Therefore, the previous orbifold $SU(3)_C\times SU(3)$ model can be embedded on the
4-brane at $z=0$. Similar to the discussions in Refs.~\cite{LL1, hn, kdw}, the tree level weak mixing angle
$\sin^2\theta_W$ at the $SU(3)$ unification
scale is 0.25, which is close to that at weak scale (0.2312). And the
correct hypercharges for the Standard Model quarks and leptons can be obtained
from the gauge invariant of the Yukawa couplings and
 four anomaly-free conditions: $[SU(3)_C]^2 U(1)_Y$, $[SU(2)_L]^2 U(1)_Y$,
$[U(1)_Y]^3$ and $[\rm{Gravity}]^2 U(1)_Y$, since we only introduce the extra exotic particles
which are vector-like under the Standard Model gauge symmetry.

For a generic multiplet $\Phi(x^{\mu}, y)$ which fills a representation of the gauge
group $SU(3)$ on the 4-brane at $z=0$,
we can define the parity operator $P^y$ 
\begin{eqnarray}
\Phi(x^{\mu},y)&\to \Phi(x^{\mu},-y )=\eta_{\Phi} P^{\l_{\Phi}}\Phi(x^{\mu},y)
(P^{-1})^{m_{\Phi}}~,~\,
\end{eqnarray}
where $\eta_{\Phi}=\pm1$.

The KK mode expansions for the bulk fields and general model
building can be found in Ref.~\cite{LTJ1}.
Choosing $\eta_{\Phi_u}=\eta_{\Phi_d}=-1$, we obtain the particle spectrum
for the vector multiplet and Higgs fields which is given in Table 1. We also
present the  gauge superfields,
the number of the 4-dimensional supersymmetry and gauge groups on the
3-branes at the fixed points or 4-branes on the fixed lines in Table 2. 
For the zero modes, the 6-dimensional $N=2$ supersymmetry and the $SU(6)$ gauge
symmetry are broken down to the 4-dimensional $N=1$ supersymmetry and
the $SU(3)_C\times SU(2)_L\times U(1)_Y\times U(1)'$ gauge symmetry.

\subsection{Anomaly Cancellation and Exotic Particles}

The anomaly from the massive KK modes of the Higgs hypermultiplets $\Psi_u$, $\Psi_d$
and $\Psi_{S_i}$ on the 4-brane at $z=0$
can be cancelled by introducing the suitable
Chern-Simons terms on the 4-brane at $z=0$ or the bulk topological 
term~\cite{anomaly, LL2}
because of the anomaly inflow~\cite{Callan, Green}. The chiral zero modes for 
$\Sigma_5^{\hat{a}, b}$ are just a pair of Higgs doublets 
with quantum number $(1; 2; 3/2; 0)$ and $(1; 2; -3/2; 0)$ under the  
$SU(3)_C\times SU(2)_L\times U(1)_Y \times U(1)'$ gauge symmetry,
then, they do not contribute to anomaly.
So, only the chiral zero modes of the Higgs hypermultiplets $\Psi_u$, $\Psi_d$
and $\Psi_{S_i}$ will contribute to the localized anomaly, which is split on
the 3-branes at $(y=0, z=0)$ and $(y=\pi R_1, z=0)$.
In fact, the 4-dimensional anomaly cancellation is sufficient to ensure the
consistency of the higher dimensional orbifold theory~\cite{anomaly, LL2}. 
In other words,
 we only need to consider the anomaly localized 
on the 3-branes at $(y=0, z=0)$ and $(y=\pi R_1, z=0)$ due to the chiral zero modes
of the Higgs on the 4-brane at $z=0$, the Standard Model
fermions and exotic particles. In addition,  
if the anomaly localized on the 3-brane at $(y=0, z=0)$
and the anomaly localized on the 3-brane at $(y=\pi R_1, z=0)$
have the opposite sign and same magnitude, the total anomalies
can be cancelled by introducing the suitable
Chern-Simons terms on the 4-brane at $z=0$ or the bulk topological 
term~\cite{anomaly, LL2} due to the anomaly inflow~\cite{Callan, Green}.
 Therefore, the anomaly free constraint is that
the sum of the localized anomalies on the
3-branes at $(y=0, z=0)$ and $(y=\pi R_1, z=0)$ should be
zero.

The anomaly from the Standard Model Higgs and fermions under the
Standard Model gauge symmetry cancel each other, and
we only introduce the extra exotic particles which are vector-like under
the Standard Model gauge symmetry. So, the anomaly will of course involve at
least one $U(1)'$.

To generate the $\mu$ term, we introduce a singlet $S$ with $U(1)'$ charge $s$.
In order to cancel the $U(1)'$ anomalies, we will introduce extra exotic 
particles\footnote{For simplicity, we do not consider the other exotic
particles, for example, the particles transforms as
$ (3; 1; 2/3; \gamma)$ and $(\bar 3; 1; -2/3; -(\gamma+2s))$
or
$ (3; 1; 2/3; \gamma)$ and $(\bar 3; 1; -2/3; -(\gamma-s))$.}:
 $k_3$ copies of $F_i$ and $\bar F_i$
where $i=1, 2, ..., k_3$,
$k_3'$ copies of $F_i'$ and $\bar F_i'$, 
$k_2$ copies of $X_i$ and $\bar X_i$,
$k_2'$ copies of $X_i'$ and $\bar X_i'$, 
$k_1$ copies of $Y_i$ and $\bar Y_i$,
$k_1'$ copies of $Y_i'$ and $\bar Y_i'$, 
$k_0$ copies of $Z_i$ and $\bar Z_i$,
$k_0'$ copies of $Z_i'$ and $\bar Z_i'$.
We also include the right handed neutrinos $\bar \nu_i$.
The quantum numbers for the Standard Model
fermions and extra exotic particles under the $SU(3)_C\times SU(2)_L\times U(1)_Y \times
U(1)'$ gauge symmetry are given in 
Table 3.
And the anomaly free conditions for $[SU(3)_C]^2 U(1)'$,
$[SU(2)_L]^2 U(1)'$, $[U(1)_Y]^2 U(1)'$, $U(1)_Y[U(1)']^2$,
$[U(1)']^3$ and $[{\rm Gravity}]^2 U(1)'$ respectively are
\begin{eqnarray}
- 3 + 2 k_3 - k_3' = 0~,~\,
\label{A1}
\end{eqnarray}
\begin{eqnarray}
 ( 2 k_2 - k_2' + 1 ) ( c_1 + c_2 ) + 3 ( a' + 3 b' ) = 0 ~,~\,
\label{A2}
\end{eqnarray}
\begin{eqnarray}
 (\frac{9}{2}  - 2 k_1 + k_1' - k_2 + \frac{k_2'}{2} - \frac{2 k_3}{3}  +
\frac{k_3'}{3} ) ( c_1 + c_2 ) + \frac{3}{2} ( a' + 3 b' ) = 0~,~\,
\label{A3}
\end{eqnarray}
\begin{eqnarray}
&& -3 a'^{2}+3 b'^{2} + 3 (a'+c_1)^2-6(b'+c_2)^2+3(b'+c_1)^2 + 2 c_2^2 - 2 c_1^2\\ \nonumber
&& +k_3((d_3+2s)^2-d_3^2) + k_3' ((d_3'-s)^2-d_3'^{2}) + k_2 (d_2^2-(d_2+2s)^2) \\ \nonumber
&& + k_2' (d_2'^{2}-(d_2'-s)^2) +k_1((d_1+2s)^2-d_1^2) + k_1' ((d_1'-s)^2-d_1'^{2})=0~,~\,
\label{A4}
\end{eqnarray}
\begin{eqnarray}
 &&6 a'^{3} - 3 ( a' + \delta )^3 - 3 ( a' + c_1 )^3 + 18 b'^{3} - 9 ( b' + c_2 )^3
- 9 ( b' + c_1 )^3  \\ \nonumber
&&+k_0 ( d_0^3 - ( d_0 + 2 s )^3 ) + k_0' (d_0'^3 - ( d_0'- s )^3 ) + k_1
( d_1^3 - ( d_1 + 2 s )^3 )  \\ \nonumber
&&+k_1' ( d_1'^3 - ( d_1' - s )^3 ) + 2 k_2 ( d_2^3 - ( d_2 + 2 s )^3 ) +
2 k_2' ( d_2'^3 - ( d_2' - s )^3 )  \\ \nonumber
&&+3 k_3 ( d_3^3 - ( d_3 + 2 s )^3 ) + 3 k_3' ( d_3'^3 - ( d_3' - s )^3)
+ c_1^3 + c_2^3 + 7 s^3 = 0 ~,~\,
\label{A5}
\end{eqnarray}
\begin{eqnarray}
(-12 + 2 k_0 - k_0' + 2 k_1 - k_1' + 4 k_2 - 2 k_2' + 6 k_3 - 3 k_3') (c_1+c_2)
+ 3(c_2-\delta)= 0 ~.~\,
\label{A6}
\end{eqnarray}
Using Eqs. (\ref{A1}) and (\ref{A2}), we can reduce the Eq. (\ref{A3})
to
\begin{eqnarray}
3 - 2 k_1 + k_1' - 2 k_2 + k_2' = 0~,~\,
\label{A3P}
\end{eqnarray}
and then, reduce Eq. (\ref{A6}) to
\begin{eqnarray}
(2 k_0 - k_0' + 2 k_2 - k_2') (c_1+c_2) + 3(c_2-\delta) = 0~.~\,
\label{A6P}
\end{eqnarray}

The simple solutions of Eqs. (\ref{A1}) and (\ref{A3P}) for $k_i$ and $k_i'$
are
\begin{eqnarray}
k_1=2~,~k_1'=1~,~ k_2=0~,~ k_2'=0~,~k_3=2~,~k_3'=1~,~  \,
\label{SSIA}
\end{eqnarray}
\begin{eqnarray}
k_1=2~,~k_1'=0~,~ k_2=0~,~ k_2'=1~,~k_3=2~,~k_3'=1 ~,~  \,
\label{SSIB}
\end{eqnarray}
\begin{eqnarray}
k_1=1~,~k_1'=1~,~ k_2=1~,~ k_2'=0~,~k_3=2~,~k_3'=1~,~  \,
\label{SSIC}
\end{eqnarray}
\begin{eqnarray}
k_1=1~,~k_1'=0~,~ k_2=1~,~ k_2'=1~,~k_3=2~,~k_3'=1 ~,~  \,
\label{SSID}
\end{eqnarray}
\begin{eqnarray}
k_1=0~,~k_1'=1~,~ k_2=2~,~ k_2'=0~,~k_3=2~,~k_3'=1~.~  \,
\label{SSIE}
\end{eqnarray}

The $U(1)'$ charge for the right handed neutrinos is
$-(a'+\delta)$, and we consider the following five scenarios for $\delta$:

\vspace{0.4cm}

(I) $\delta=c_2$ and $ a'+c_2-s=0$.

\vspace{0.4cm}

(II) $\delta = c_2+s$ and $2 a'+2c_2+3s=0$.

\vspace{0.4cm}

(III) $\delta = c_2+3s$ and $a'+c_2+2s=0$.

\vspace{0.4cm}

(IV) $\delta = c_2-s$ and $a'+c_2-2s=0$.

\vspace{0.4cm}

(V) $\delta = c_2-2s$ and $a'+c_2-3s=0$.

\vspace{0.4cm}

It is not difficult to find the irrational solutions to the
Eqs. (\ref{A1}-\ref{A6}).
Mathematically speaking, the rational $U(1)'$ charges for all the particles
are equivalent to the integer $U(1)'$ charges for all the particles.
To be concrete, we present
the sample rational solutions to the 
Eqs. (\ref{A1}-\ref{A6})
 in Tables 4, and 5 for above five scenarios, 
with the $k_1$, $k_1'$, $k_2$, $k_2'$,
$k_3$ and $k_3'$ given in 
 Eqs.  (\ref{SSIB}), and
(\ref{SSIC}),  respectively.
If $k_0=0$ or $k_0'=0$, we do not have the exotic particles
$Z_i$ and $\bar Z_i$ or $Z'_i$ and $\bar Z'_i$. Thus,
$d_0$ or $d'_0$ is not relevant. For simplicity,
we write $d_0=X$ or $d'_0=X$ if $k_0=0$ or $k_0'=0$ in the Tables.

\subsection{Gauge Coupling Unification}

For the gauge coupling unification, the known problem is that the localized
gauge kinetic terms on the 3-branes at the fixed points or on the
4-branes at the fixed lines may spoil the desired predictivity
of higher dimensional gauge theory. For example, as shown in the
Table 2, the  smaller gauge symmetries other than $SU(6)$
are respected on some 3-branes or 4-branes. 
This implies that at the fundamental scale, or 
cutoff scale $M_*$ in the theory, the
$SU(6)$ gauge coupling relation might not be preserved to a good
precision. In this paper, we consider the $SU(6)$ unification scale
as the fundamental scale, or 
cutoff scale $M_*$ in the theory.
To suppress these uncertainties, one may assume that 
the theory is strongly coupled at the cutoff scale, and the gauge couplings
for the bulk and localized kinetic terms are of the same magnitude
at the cutoff scale. If the cutoff scale is much higher than the
compactification scale, the uncertainties from the localized terms are then
suppressed to a good precision~\cite{hall}. We are not going to pursue this
idea in this paper. Basically it gives a strong constraint. On the
other hand, we would like to study by numerical calculations how well
the gauge coupling unification can be achieved using the particle contents
involved and leave the problem for further study (for example from the
point of view of string theory). In fact, at tree-level, there is no localized
gauge kinetic term on the orbifold fixed point in the weakly coupled
heterotic string theory.

As an example, we discuss the scenario II given in Table 4.
Clearly, it's the KK modes of the Higgs multiplets on the 
4-brane at $z=0$ that contribute to the asymmetric radiative corrections to the
$SU(3)_C$ and $SU(3)$ gauge coupling runnings. 
One can see from Table 1 that the even modes along the $z$ direction 
$V^{a,b}$, $V^{\hat{a},b}$, $\Sigma_5^{a,b}$ and $\Sigma_5^{\hat{a},b}$,
which are relevant for gauge coupling runnings at the energy scale $1/R_1 < \mu < 1/R_2$, form 
adjoint representations under the $SU(3)_C\times SU(3)$ gauge symmetry. For energy
scale above $1/R_2$, the spectrum of massive KK modes is actually
$SU(6)$ symmetric in the gauge sector.
One can also see from the Table 1 that
the  Higgs triplets give no asymmetric radiative corrections to the
$SU(2)_L$ and $U(1)_Y$ gauge coupling runnings.
Unlike the story in Ref. \cite{LL1}, the massive KK modes do not
help us much on the unification of three gauge couplings. Interestingly,
the extra exotic particles can indeed do the job. Notice that
there are quite a few exotic particles which are $SU(2)_L$ singlets
and charged under $U(1)_Y$,
the exotic particles do accelerate the $SU(3)$ asymmetric logarithmic gauge coupling 
runnings. For the exotic particles with masses around $1.2$ TeV, we can have
the gauge coupling unification to a good precision (less than $1\%$)
if $1/R_1 \simeq 3-4$ TeV
and $1/R_2 \simeq 6/R_1$ and $M_* \simeq 50/R_1$. The gauge coupling ($\alpha$)
at the unification scale $M_*$ is around $1/55$, which implies that the gauge coupling 
$\alpha_1^{\prime}$ for $U(1)'$
is about $1/105$ at energy  scale $1.2$ TeV.

The discussions of gauge coupling unification
for all five scenarios in Tables 4-5 are similar. In short,
the gauge coupling unification can be achieved at the $100\sim 200$ TeV
if the compactification scale for the fifth dimension ($1/R_1$)
is about $3\sim 4$ TeV.

\subsection { $SU(3)_C\times SU(2)_L\times U(1)_Y \times U(1)'$ 
Gauge Symmetry Breaking and Particle Masses}

In order to avoid the proton decay problem from the non-renormalizable
operators, we put the Standard Model quarks, $Q_i$, $\bar U_i$ and $\bar D_i$,
and the exotic triplet particles $F_i$, $\bar F_i$, $F_i'$ and $\bar F_i'$
on the 3-brane at $(y=0, z=0)$.  Moreover, we put the Standard Model
leptons $L_i$, $\bar E_i$ and $\bar \nu_i$, $S$, and the rest
of the exotic particles, $X_i$, $\bar X_i$, $X_i'$, $\bar X_i'$,
$Y_i$, $\bar Y_i$, $Y_i'$, $\bar Y_i'$, $Z_i$, $\bar Z_i$,
$Z_i'$ and $\bar Z_i'$ on the 3-brane at $(y=\pi R_1, z=0)$.
As an example, we consider the scenario II in Table 4 in the last subsection, 
the proton decay operators are suppressed by the factor
$e^{-3(\pi R_1 M_* )^2/4} \sim e^{-18506} \sim 10^{-8142}$~\cite{NAMS}.

First, let us discuss the $SU(3)_C\times SU(2)_L\times U(1)_Y \times U(1)'$
gauge symmetry breaking and briefly
review the results in Ref.~\cite{JPT}.

Because on the 4-brane at $z=0$, there exists the 4-dimensional $N=2$
supersymmetry, we have to put the superpotential on the
3-brane at $(y=\pi R_1, z=0)$ where only the 4-dimensional $N=1$
supersymmetry is preserved. The localized superpotential
for the Higgs fields are
\begin{eqnarray}
S=\int d^4x dy dz \delta(y-\pi R_1) \delta(z) \left[\int d^2\theta
({h^5} S H_d H_u + \lambda^5 S_1 S_2 S_3) + H. C. \right]
~,~\,
\label{SSSS}
\end{eqnarray}
where $h^5$ and $\lambda^5$ are 5-dimensional couplings. Similar notations are
used for the other 5-dimensional couplings.

In the 4-dimensional effective theory, we obtain the
superpotential
\begin{eqnarray}
W &=& h S H_d H_u + \lambda S_1 S_2 S_3 ~,~\,
\end{eqnarray} 
where $h=h^5/(\pi R_1 M_*)$ and 
$\lambda = \lambda^5/(\pi R_1 M_*)^{3/2}$,
because there is a normalization
factor $1/(\pi R_1 M_*)^{1/2}$ for each field on the 4-brane at $z=0$~\cite{HMR}.
For the  scenario II in Table 4 in the last subsection,
 $(\pi R_1 M_*)^{1/2}\sim 12.5$.
Because it is natural to take $h^5\sim \lambda^5$,  we obtain 
$h \sim 10 \lambda$ which is crucial to generate the
$Z-Z'$ mass hierarchy.

The corresponding $F$-term scalar potential is 
\begin{eqnarray}
V_F &=& h^2 \left( |H_d|^2 |H_u|^2 + |S|^2 |H_d|^2 + |S|^2|H_u|^2\right)
\nonumber\\&&
+\lambda^2 \left(|S_1|^2 |S_2|^2 + |S_2|^2 |S_3|^2 + |S_3|^2 |S_1|^2\right)
~.~\,
\end{eqnarray} 
The $D$-term scalar potential is 
\begin{eqnarray}
V_D &=& {{G^2}\over 8} \left(|H_u|^2 - |H_d|^2\right)^2 
\nonumber\\&&
+{1\over 2} g_{Z'}^2\left( s |S|^2 +  
c_1 |H_d|^2 + c_2 |H_u|^2 -s (|S_1|^2 +|S_2|^2)+ 2 s |S_3|^2\right)^2 ~,~\,
\end{eqnarray}  
where $G^2=g_1^{2} +g_2^2$; $g_1, g_2$,  and $g_{Z'}$ are the coupling constants for
$U(1)_Y, SU(2)_L$ and $U(1)^{\prime}$, respectively.

In addition, we introduce the supersymmetry breaking soft terms
\begin{eqnarray}
V_{soft}^{(a)} &=& m_{H_d}^2 |H_d|^2 + m_{H_u}^2 |H_u|^2 + m_S^2 |S|^2 +
\sum_{i=1}^3 m_{S_i}^2 |S_i|^2
\nonumber\\&&
-(A_h h S H_d H_u + A_{\lambda} \lambda S_1 S_2 S_3 
+m_{S S_1}^2 S S_1 + m_{S S_2}^2 S S_2
+ {\rm H. C.})~.~\,  
\label{vsoft}
\end{eqnarray} 

To eliminate the runaway directions of the potential that are not bounded from
below, we
require
\begin{eqnarray}
m_S^2 + m_{S_1}^2 +2 m_{S S_1}^2 > 0 ~,~ m_S^2 + m_{S_2}^2 +2 m_{S S_2}^2 > 0
~.~\,
\end{eqnarray} 
The first condition avoid the runaway direction in which $<S> = <S_1>$ with
the other VEVs vanishing, for which the quartic and cubic terms in the
potential are flat. Similarly for the second condition.

The $Z-Z'$ mass matrix is
\begin{eqnarray}
M_{Z-Z'} =\left(\matrix{M_{Z}^2 & M_{Z Z'}^2\cr
M_{Z Z'}^2 &  M_{Z'}^2\cr}\right) ~,~ \,
\end{eqnarray}
where
\begin{eqnarray}
M_{Z}^2 = {G^2\over 2} (v_1^2 +v_2^2)
~,~\,
\end{eqnarray} 
\begin{eqnarray}
M_{ Z'}^2 = 2 g_{Z'}^2 
\left(  Q_S^2 <S>^2 + Q_{H_1}^2 v_1^2 + Q_{H_2}^2 v_2^2 + \sum_{i=1}^3 Q_{S_i}^2 <S_i>^2
\right)~,~\,
\end{eqnarray} 
\begin{eqnarray}
M_{Z Z'}^2 = g_{Z'} G ( Q_{H_1} v_1^2 - Q_{H_2} v_2^2)
~,~\,
\end{eqnarray} 
where 
\begin{eqnarray}
\langle H_1^0 \rangle \equiv v_1~,~ \langle H_2^0\rangle
 \equiv v_2~.~\,
\end{eqnarray}   
The mass eigenvalues are
\begin{eqnarray}
M_{Z_1, Z_2}^2 = {1\over 2} \left(M_Z^2 + M_{Z'}^2 \mp 
\sqrt {(M_Z^2-M_{Z'}^2)^2 + 4 M_{Z Z'}^4 } \right) 
~,~\,
\end{eqnarray} 
and the $Z-Z'$ mixing angle $\theta_{Z-Z'}$ is given by
\begin{eqnarray}
\theta_{Z-Z'} = {1\over 2} {\rm arctan} \left({{2 M_{ZZ'}^2}
\over\displaystyle {M_{Z'}^2 - M_Z^2}}\right)
~,~\,
\end{eqnarray} 
which is constrained to be less than a few times $10^{-3}$.

It has been shown that this potential can generate 1 TeV scale VEVs
for $S_1$, $S_2$ and $S_3$, and $100\sim 200$ GeV scale VEVs
for $S$, $H_d$ and $H_u$~\cite{JPT}. 
As an example, we discuss the potential with the $U(1)'$ charges for the Higgs
 and the low energy $U(1)'$ gauge coupling 
in the scenario II given in Table 4 in
the last subsection.
With the input
$h=0.75$, $\lambda = 0.075$
$A_h = A_{\lambda} =1.0,
m_{H_d}^2 = m_{H_u}^2= m_{S}^2 = -0.010,
m_{S_1}^2 = m_{S_2}^2=0.031, m_{S_3}^2 = -0.010,
m_{S S_1}^2= m_{S S_2}^2=-0.010, c_1=1/6,
c_2=5/6$,  and $s=-1$, we obtain that
the VEVs for Higgs fields at the  minimum are 
$<H_d> =0.626, <H_u>=0.631,
<S>=0.949, <S_1>=6.45, <S_2>=6.44$, and $<S_3>=6.43$.
The input parameters with dimensions of mass or mass-squared are chosen
in arbitrary units. After finding an acceptable minimum, they are rescaled
so that $\sqrt{v_1^2+v_2^2}\simeq 174$ GeV. After rescaling, the 
VEVs for  Higgs fields at the minimum are 
$<H_d>=122.5$ GeV, $<H_u>=123.5$ GeV,
$<S>=185.9$ GeV, $<S_1>=1262.8$ GeV, $<S_2>=1260.2$ GeV, and $<S_3>=1258.5$ GeV.
Thus, we obtain that $\mu= h<S>=139.4$ GeV, $M_Z=91.37$ GeV,
$M_{Z'}=1513$ GeV and $\theta_{Z-Z'}= 1.15\times 10^{-3}$.

In short, we can solve the $\mu$ problem and generate $Z-Z'$ mass hierarchy
naturally.

Second, we discuss the Yukawa couplings for the 
Standard Model quarks and leptons, and the exotic particle
masses. We can introduce the following superpotentials localized on the 3-branes
at $(y=0, z=0)$ and $(y=\pi R_1, z=0)$
\begin{eqnarray}
S&=&\int d^4x dy dz \delta(y) \delta(z) \int d^2\theta
\left(
  y_{uij}^5 Q_i H_u {\bar U_j}+  y_{dij}^5 Q_i H_d {\bar D_j}
\right.\nonumber\\&&\left. 
+ h_{Fi}^5 S_3 F_i {\bar F_i} +  h^{\prime5}_{Fi} S_1 F_i' {\bar F_i'}
+{\lambda}^{\prime5}_{Fi} S_2 F_i' {\bar F_i'}
 + H. C. \right)
\nonumber\\&&
+ \int d^4x dy dz \delta(y-\pi R_1) \delta(z) \int d^2\theta
\left( y_{eij}^5 L_i H_d {\bar E_j}
\right.\nonumber\\&&\left.
+ h_{Xi}^5 S_3 X_i {\bar X_i} +  h^{\prime 5}_{Xi} S_1 X_i' {\bar X_i'}
+{\lambda}^{\prime5}_{Xi} S_2 X_i' {\bar X_i'}
\right.\nonumber\\&&\left.
+ h_{Yi}^5 S_3 Y_i {\bar Y_i} +  h^{\prime5}_{Yi} S_1 Y_i' {\bar Y_i'}
+{\lambda}^{\prime5}_{Yi} S_2 Y_i' {\bar Y_i'}
\right.\nonumber\\&&\left.
+ h_{Zi}^5 S_3 Z_i {\bar Z_i} +  h^{\prime5}_{Zi} S_1 Z_i' {\bar Z_i'}
+{\lambda}^{\prime5}_{Zi} S_2 Z_i' {\bar Z_i'}
 + H. C. \right)
~.~\,
\end{eqnarray}  
In the 4-dimensional effective theory, we have
the following superpotential
\begin{eqnarray}
W&=& y_{uij} Q_i H_u {\bar U_j} +  y_{dij} Q_i H_d {\bar D_j}
+ y_{eij} L_i H_d {\bar E_j} 
\nonumber\\&&
+ h_{Fi} S_3 F_i {\bar F_i} +  h^{\prime}_{Fi} S_1 F_i' {\bar F_i'}
+{\lambda}^{\prime}_{Fi} S_2 F_i' {\bar F_i'}
\nonumber\\&&
+ h_{Xi} S_3 X_i {\bar X_i} +  h^{\prime}_{Xi} S_1 X_i' {\bar X_i'}
+{\lambda}^{\prime}_{Xi} S_2 X_i' {\bar X_i'}
\nonumber\\&&
+ h_{Yi} S_3 Y_i {\bar Y_i} +  h^{\prime}_{Yi} S_1 Y_i' {\bar Y_i'}
+{\lambda}^{\prime}_{Yi} S_2 Y_i' {\bar Y_i'}
\nonumber\\&&
+ h_{Zi} S_3 Z_i {\bar Z_i} +  h^{\prime}_{Zi} S_1 Z_i' {\bar Z_i'}
+{\lambda}^{\prime}_{Zi} S_2 Z_i' {\bar Z_i'}
~.~\,
\end{eqnarray}  
Therefore, all the exotic particles can obtain the masses around 1 TeV
after the gauge symmetry breaking because the VEVs for $S_i$ are at the order of 1 TeV.

Third, one might notice that there are zero modes for $S_u^c$, $S_d^c$, and 
$\Sigma_5^{\hat{a}, b}$ from Table 1. The zero modes of
 $S_u^c$ and $S_d^c$ can obtain the masses
via the following localized superpotential
\begin{eqnarray}
S &=& \int d^4x dy dz \delta(y) \delta(z) \left[\int d^2\theta
({\lambda_{S1}^5} S_1 S_u^c S_d^c + {\lambda_{S2}^5} S_2 S_u^c S_d^c) + H. C. \right]
\nonumber\\&&
+ \int d^4x dy dz \delta(y-\pi R_1) \delta(z) \left[\int d^2\theta
({\lambda_{S1}^{\prime 5}} S_1 S_u^c S_d^c + {\lambda_{S2}^{\prime 5}} S_2 S_u^c S_d^c) + H. C. \right],~\,
\label{SuSd}
\end{eqnarray}
or in the 4-dimensional effective theory,  the superpotential is
\begin{eqnarray}
W &=& ({\lambda_{S1}}+{\lambda_{S1}^{\prime}})  S_1 S_u^c S_d^c + 
({\lambda_{S2}}+{\lambda_{S2}^{\prime }})  S_2 S_u^c S_d^c
~.~\,
\end{eqnarray}
Thus, the $S_u^c$ and $S_d^c$ can have masses at the order of 1 TeV
if $\lambda_{S1}$, $\lambda_{S1}^{\prime}$, $\lambda_{S2}$
and $\lambda_{S2}^{\prime}$ are at the order of 1.

Moreover, $\Sigma_5^{\hat{a}, b}$ are just one pair of Higgs doublets 
with quantum number $(1; 2; 3/2; 0)$ and $(1; 2; -3/2; 0)$ under the 
$SU(3)_C\times SU(2)_L\times U(1)_Y \times U(1)'$ gauge symmetry.
 The fermionic components of $\Sigma_5^{\hat{a}, b}$ can obtain the gaugino masses
 via the supersymmetry breaking, for instance, the gaugino masses
are proportional to $1/R_1$ in the Scherk-Schwarz
mechanism supersymmetry breaking. Meanwhile, the 
bosonic components of $\Sigma_5^{\hat{a}, b}$ receive the masses of order
of $1/(4\pi R_1)$ through the radiative corrections. Because $1/(4\pi R_1)$ is
about several hundreds of GeV in our model, the reasonable masses for $\Sigma_5^{\hat{a}, b}$
are around several hundreds of GeV range. Alternatively, $\Sigma_5^{\hat{a}, b}$
can obtain masses at the order of several GeVs
 via the non-renormalizable localized superpotential,
\begin{eqnarray}
S&=&\int d^4x dy dz \delta(y-\pi R_1) \delta(z) \left[\int d^2\theta
{{\lambda_{SS1}^{5}} \over M_*} S S_1 tr(\Sigma_5^{\hat{a}, b} \Sigma_5^{\hat{a}, b})  + H. C. \right]
~,~\,
\end{eqnarray}
or in the 4-dimensional effective theory, the superpotential is
\begin{eqnarray}
W&=& {{\lambda_{SS1}} \over M_*} S S_1 tr(\Sigma_5^{\hat{a}, b} \Sigma_5^{\hat{a}, b})
~.~\,
\end{eqnarray}
We also want to point out that
$\Sigma_5^{\hat{a}, b}$ can not couple to the Standard Model fermions and
exotic particles due to the $U(1)'$ gauge symmetry, so, we have no low bounds
on the masses of $\Sigma_5^{\hat{a}, b}$, as long as they are not massless which may cause problem
in cosmology.

Fourth, 
the dimension-5 operators  $y_{\nu ij} L_i L_j H_u H_u/M_*$ are forbidden by
 $U(1)'$ gauge symmetry and we would like to
 discuss the neutrino masses for all five scenarios with different $\delta$. 
 For scenario (I),
we will give a complete discussion, and for the other scenarios, we only give 
brief descriptions. 

(I) $\delta=c_2$ and $ a'+c_2-s=0$.
The neutrinos can obtain masses via the localized superpotential
\begin{eqnarray}
S&=&\int d^4x dy dz \delta(y-\pi R_1) \delta(z) \left[\int d^2\theta
h_{rij}^5 S_3 {\bar \nu_i} {\bar \nu_j} + y_{\nu ij}^5 H_u L_i {\bar \nu_j}  + H. C. \right]
~,~\,
\end{eqnarray}
or in the 4-dimensional effective theory, the superpotential is
\begin{eqnarray}
W&=& h_{rij} S_3 {\bar \nu_i} {\bar \nu_j} + y_{\nu ij} H_u L_i {\bar \nu_j} 
~.~\,
\end{eqnarray}
In the basis $\{\nu_1, \nu_2, \nu_3, {\bar \nu_1}, {\bar \nu_2}, {\bar \nu_3} \}$,
the neutrino mass matrix is
\begin{eqnarray}
M = \left(\matrix{ 0 & M_D  \cr
M_D^T & M_N \cr}\right)
~,~ \,
\end{eqnarray}
where 
\begin{eqnarray}
M_D  = \left(\matrix{ y_{\nu 11} <H_u^0>  & y_{\nu 12} <H_u^0>  & y_{\nu 13} <H_u^0> \cr
 y_{\nu 21} <H_u^0>  & y_{\nu 22} <H_u^0>  & y_{\nu 23} <H_u^0>   \cr
 y_{\nu 31} <H_u^0>  & y_{\nu 32} <H_u^0>  & y_{\nu 33} <H_u^0>  \cr}\right)~,~ \,
\end{eqnarray}
\begin{eqnarray}
M_N  = \left(\matrix{ h_{r11} <S_3>  &  h_{r12} <S_3>/2 & h_{r13} <S_3> /2 \cr
 h_{r12} <S_3>/2  &  h_{r22} <S_3>  & h_{r23} <S_3> /2     \cr
 h_{r13} <S_3> /2 &  h_{r23} <S_3> /2 &   h_{r33} <S_3> \cr}\right)~.~ \,
\end{eqnarray}
The left-handed neutrino mass matrix $M_{\nu}$ is equivalent to
\begin{eqnarray}
M_{\nu}=M_D M_N^{-1} M_D^T ~.~ \,
\end{eqnarray}
Because $<S_3>$ is at the order of 1 TeV in our model,  the 
right-handed neutrino masses are about 1 TeV if $h_{rij}$ is
at the order of 1. Therefore, in order to have the realistic
active neutrino masses, we obtain that the neutrino Dirac masses should be
at the order of $10^{-3}$ GeV, {\it i.e.}, the Yukawa couplings
($y_{\nu ij}$)  should
be at the order of $10^{-5}\sim 10^{-6}$ or less.

(II) $\delta = c_2+s$ and $2 a'+2c_2+3s=0$.
The neutrinos can obtain masses via the following superpotential in the 4-dimensional
effective theory
\begin{eqnarray}
W&=& h_{rij} S_1 {\bar \nu_i} {\bar \nu_j}
+  h'_{rij} S_2 {\bar \nu_i} {\bar \nu_j}
 + y_{\nu ij}  {S\over M_*} H_u L_i {\bar \nu_j} 
~.~\,
\end{eqnarray}
Because in our model, $<S_1>$ and $<S_2>$ are at the order of 1 TeV, the 
right-handed neutrino masses can be about 5 TeV if $h_{rij}$ and $h'_{rij}$ are
about 2.5.
So, in order to have the realistic
active neutrino masses, we obtain that the couplings $y_{\nu ij}$ for the non-renormalizable
terms should be at the order of or smaller than 0.1 because $M_* \sim 2\times 10^5$ GeV.

(III) $\delta = c_2+3s$ and $a'+c_2+2s=0$.
The neutrinos can obtain masses via the following superpotential in the 4-dimensional
effective theory
\begin{eqnarray}
W&=& h_{rij} S_3 {\bar \nu_i} {\bar \nu_j}
 + y_{\nu ij}  {1 \over {(\pi R_1 M_*)^{1/2}}} {{S S_3}\over M_*^2} H_u L_i {\bar \nu_j} 
~,~\,
\end{eqnarray}
where compare to the tree-level Yukawa coupling,
we have an extra $1/(\pi R_1 M_*)^{1/2}$ factor because $S_3$ is on the 4-brane
at $z=0$, and then, there is a normalization factor $1/(\pi R_1 M_*)^{1/2}$ for it~\cite{HMR}.
The right-handed neutrino masses are about 1 TeV if $h_{rij}$ is
at the order of 1.
So, in order to have the realistic
active neutrino masses, we obtain that the couplings $y_{\nu ij}$ for the non-renormalizable
terms should be at the order of or smaller than 10.

(IV) $\delta = c_2-s$ and $a'+c_2-2s=0$.
The neutrinos can obtain masses via the following superpotential in the 4-dimensional
effective theory
\begin{eqnarray}
W&=& h_{rij} S_3 {\bar \nu_i} {\bar \nu_j}
 +  {1 \over {(\pi R_1 M_*)^{1/2}}} 
(y_{\nu ij} {{S_1}\over M_*} + y'_{\nu ij} {{S_2}\over M_*}) H_u L_i {\bar \nu_j}
~.~\,
\end{eqnarray}
The right-handed neutrino masses are about 1 TeV if $h_{rij}$ is
at the order of 1.
So, in order to have the realistic
active neutrino masses, we obtain that the couplings $y_{\nu ij}$ and $y'_{\nu ij}$ for the non-renormalizable
terms should be at the order of or smaller than $10^{-2}$. If there is cancellation 
in $(y_{\nu ij}  <S_1> + y'_{\nu ij} <S_2>)$,
the couplings $y_{\nu ij}$ and $y'_{\nu ij}$ can be at the order of 0.1.

(V) $\delta = c_2-2s$ and $a'+c_2-3s=0$.
The neutrinos can obtain masses via the following superpotential in the 4-dimensional
effective theory
\begin{eqnarray}
W&=& h_{rij} S_3 {\bar \nu_i} {\bar \nu_j}
 + y_{\nu ij} {1 \over {\pi R_1 M_*}} {{S_1 S_2}\over M_*^2} H_u L_i {\bar \nu_j}  
~.~\,
\end{eqnarray}
The right-handed neutrino masses are about 1 TeV if $h_{rij}$ is
at the order of 1.
So, in order to have the realistic
active neutrino masses, we obtain that the couplings $y_{\nu ij}$ for the non-renormalizable
terms should be at the order of or smaller than $10$.

In short, the very tiny realistic neutrino masses can
be generated naturally in the scenarios (II), (III), (IV) and (V).

\section{$SU(6)$ Model on $M^4 \times T^2/(Z_2)^4$}

In this section, we would like to discuss the 6-dimensional $N=2$ supersymmetric
$SU(6)$ model on the space-time $M^4 \times T^2/(Z_2)^4$.
Here again, on the 4-brane at $z=0$, because of the orbifold projections,
there exist only the 
$SU(3)_C\times SU(3) \times U(1)'$ gauge symmetry
and 4-dimensional $N=2$ supersymmetry, where the 
previous orbifold $SU(3)_C\times SU(3)$ model in Refs.~\cite{LL1, hn, kdw} can be embedded.

\subsection{Orbifold $SU(6)$ Breaking, and Particle Spectrum for Gauge and Higgs Fields}

In our model, there exist only the 
$SU(3)_C\times SU(3) \times U(1)'$ gauge symmetry
and 4-dimensional $N=2$ supersymmetry on the 4-brane at $z=0$, and the lepton fields
forming  triplets are on the 3-brane at $(y=\pi R_1/2, z=0)$ in which 
the $SU(3)_C\times SU(3) \times U(1)'$ gauge symmetry is preserved. Therefore,
in order to have the realistic lepton masses,
we put one pair of Higgs sextet $\Psi_u$ and $\Psi_d$ which transform as
$(1, \bar 6, c_2)$ and $(1, 6, c_1)$ under the $SU(3)_C\times SU(3) \times U(1)'$
gauge symmetry on the 4-brane at $z=0$~\footnote{
If we put one pair of the Higgs triplets on the 4-brane at $z=0$, in order to
obtain the realistic lepton masses,  we have to put the quarks and leptons on the
3-brane at $(y=0, z=0)$. The discussions for the model building are similar to those
in Section 3. However, we might not forbid the non-renormalizable
proton decay operators.}.
We require that $c_1\not=0$, $c_2\not=0$ and $c_1+c_2\not=0$. 
We also add three $SU(3)_C\times SU(3)$ singlets,  $\Psi_{S_1}$, 
$\Psi_{S_2}$ and $\Psi_{S_3}$
with $U(1)'$ charges  $-s$, $-s$ and $2s$ respectively
on the 4-brane at $z=0$, where $s\equiv-c_1-c_2$. 
In terms of the 4-dimensional $N=1$ supersymmetry language, the hypermultiplets
$\Psi_u$ and $\Psi_d$ can be decomposed into two pairs of chiral multiplets,
$(\Phi_u, \Phi_u^c)$ and $(\Phi_d, \Phi_d^c)$, and
the hypermultiplet $\Psi_{S_i}$ can be decomposed into one pair
of chiral multiplets $(S_i, S^c_i)$ in which $i=1, 2, 3$. Here, the superscript
$c$ means the charge conjugation.
To be explicit, we write down the component of $\Phi_u$ and $\Phi_d$,
\begin{eqnarray}
\Phi_u  = \left(\matrix{ \eta_u^{--} & \eta_u^{-}/{\sqrt 2}   & H_u^0/{\sqrt 2}   \cr
\eta_u^{-}/{\sqrt 2} & \eta_u^0  & H_u^+/{\sqrt 2}  \cr
H_u^0/{\sqrt 2} &  H_u^+/{\sqrt 2}  & \eta_u^{\prime ++}  \cr}\right)
~,~ \,
\end{eqnarray}
\begin{eqnarray}
\Phi_d  = \left(\matrix{ \eta_d^{++} & \eta_d^{+}/{\sqrt 2}   & H_d^0/{\sqrt 2}   \cr
\eta_d^{+}/{\sqrt 2} & \eta_d^0  & H_d^-/{\sqrt 2}  \cr
H_d^0/{\sqrt 2} &  H_d^-/{\sqrt 2}  & \eta_d^{\prime --}  \cr}\right)
~.~ \,
\end{eqnarray}
To simplify the notation, we denote the $H_u^0$ and $H_u^+$
components of $\Phi_u$ as the
Higgs doublet $H_u$, and the remaining components 
( $\eta_u^{--}$,
$\eta_u^{-}$, $ \eta_u^0$ and $\eta_u^{\prime ++}$)
of  $\Phi_u$ as $\eta_u$. Similar notation is used for
$\Phi_d$, which decomposes into $H_d$ and $\eta_d$.

In addition, we introduce extra exotic particles to cancel the anomalies, and  
put the Standard Model fermions and extra exotic particles on
the 3-branes at the orbifold fixed points, as explained later.

We choose the following matrix representations for the parity operators
$P^y$, $P^{y'}$, $P^{z}$ and $P^{z'}$, 
which are expressed in the adjoint representation of SU(6),
\begin{equation}
P^y={\rm diag}(+1, +1, +1, -1, -1, +1) ~,~ P^{y'}={\rm diag}(+1, +1, +1, +1, +1, +1)~,~\,
\end{equation}
\begin{equation}
P^z={\rm diag}(+1, +1, +1, -1, -1, -1)
~,~ P^{z'}={\rm diag}(+1, +1, +1, +1, +1, +1)~.~\,
\end{equation}

Therefore, under $P^{y}$ and $P^{z}$ parities,
the $SU(6)$ gauge generators $T^A$, where A=1, 2, ..., 35 for $SU(6)$,
are separated into four sets: $T^{a, b}$, $T^{ a, \hat b}$,
$T^{\hat a, b}$, and $T^{\hat a, \hat b}$. And
under $P^{y}$, $P^{z}$, $P^{y'}$ and $P^{z'}$, the gauge generators transform as
\begin{equation}
P^{y}~T^{a, B}~(P^{y})^{-1}= T^{a, B} ~,~ 
P^{y}~T^{\hat a, B}~(P^{y})^{-1}= - T^{\hat a, B}
~,~\,
\end{equation}
\begin{equation}
P^{z}~T^{A, b}~(P^{z})^{-1}= T^{A, b} ~,~ P^{z}~T^{A, \hat b}~(P^{z})^{-1}= - T^{A, \hat b}
~,~\,
\end{equation}
\begin{equation}
P^{y'}~T^{A, B}~(P^{y'})^{-1}= T^{A, B} ~,~ P^{z'}~T^{A, B}~(P^{z'})^{-1}= T^{A, B} 
~.~\,
\end{equation}

Furthermore, only the $SU(6)/P^z=SU(3)_C\times SU(3) \times U(1)'$
gauge symmetry and 4-dimensional $N=2$ supersymmetry are preserved on the 4-brane at $z=0$.
Therefore, the previous orbifold $SU(3)_C\times SU(3)$ model can be embedded on the
4-brane at $z=0$. Following the discussions in Refs.~\cite{LL1, hn, kdw}, the tree level weak mixing angle
$\sin^2\theta_W$ at the $SU(3)$ unification
scale is 0.25, which is close to that at weak scale. And the
correct hypercharges for the Standard Model quarks and leptons can be obtained
from the gauge invariant of the Yukawa couplings and
 four anomaly-free conditions: $[SU(3)_C]^2 U(1)_Y$, $[SU(2)_L]^2 U(1)_Y$,
$[U(1)_Y]^3$ and $[\rm {Gravity}]^2 U(1)_Y$, because we only introduce the extra exotic particles
which are vector-like under the Standard Model gauge symmetry.

For a generic multiplet $\Phi(x^{\mu}, y)$ which fills a representation of the gauge
group $SU(3)$ on the 4-brane at $z=0$,
we can define two parity operators $P^y$ and $P^{y'}$
\begin{eqnarray}
\Phi(x^{\mu},y)&\to \Phi(x^{\mu},-y )=\eta_{\Phi} P^{\l_{\Phi}}\Phi(x^{\mu},y)
(P^{-1})^{m_{\Phi}}~,~\,
\end{eqnarray}
\begin{eqnarray}
\Phi(x^{\mu},y')&\to \Phi(x^{\mu},-y' )=\eta'_{\Phi} P^{\l_{\Phi}}\Phi(x^{\mu},y')
(P^{-1})^{m_{\Phi}}~,~\,
\end{eqnarray}
where $\eta_{\Phi}=\pm1$ and $\eta'_{\Phi}=\pm1$.

The KK mode expansions for the bulk fields and the general
model discussions can be found in Ref.~\cite{LTJ1}.
Choosing $\eta_{\Phi_u}=\eta_{\Phi_d}=-1$ and $\eta'_{\Phi_u}=\eta'_{\Phi_d}=+1$,
 we obtain the particle spectrum
for the vector multiplet and Higgs fields which is given in Table 6. We also
present the  gauge superfields,
the number of the 4-dimensional supersymmetry and gauge groups on the
3-brane at the fixed points or on the 4-branes on the fixed lines in Table 7. 
For the zero modes, the 6-dimensional $N=2$ supersymmetry and the $SU(6)$ gauge
symmetry are broken down to the 4-dimensional $N=1$ supersymmetry and
the $SU(3)_C\times SU(2)_L\times U(1)_Y\times U(1)'$ gauge symmetry.

\subsection{Anomaly Cancellation and Exotic Particles}

The anomaly from the massive KK modes of the Higgs hypermultiplets $\Psi_u$, $\Psi_d$, 
and $\Psi_{S_i}$ on the 4-brane at $z=0$
can also be cancelled by introducing the suitable
Chern-Simons terms on the 4-brane at $z=0$ or the bulk topological 
term~\cite{anomaly, LL2} because of the anomaly inflow~\cite{Callan, Green}. 
And only the chiral zero modes of the Higgs hypermultiplets $\Psi_u$, $\Psi_d$, 
and $\Psi_{S_i}$ will contribute to the localized anomaly, which is split on
the 3-branes at $(y=0, z=0)$ and $(y=\pi R_1/2, z=0)$.

Because the gauge symmetry on the 3-brane at $(y=0, z=0)$ is
$SU(3)_C\times SU(2)_L \times U(1)_Y \times U(1)'$, and the gauge
symmetry on the 3-brane at $(y=\pi R_1/2, z=0)$ is
$SU(3)_C\times SU(3) \times U(1)'$,  the
particles we put on the 3-brane at $(y=\pi R_1/2, z=0)$ must
form the complete representaions under the $SU(3)_C\times SU(3) \times U(1)'$
gauge symmetry. Similarly, the 4-dimensional anomaly cancellation is 
sufficient to ensure the
consistency of the higher dimensional orbifold theory~\cite{anomaly, LL2}. In other words,
if the anomaly localized on the 3-brane at $(y=0, z=0)$
and the anomaly localized on the 3-brane at $(y=\pi R_1/2, z=0)$
for the gauge group $SU(3)_C\times SU(2)_L\times U(1)_Y \times U(1)'$
have the opposite sign and same magnitude, the total anomaly
can be cancelled by introducing the suitable
Chern-Simons terms on the 4-brane at $z=0$
or the bulk topological terms
as long as the particles
on the 3-brane at $(y=\pi R_1/2, z=0)$
form the complete representations under the $SU(3)_C\times SU(3) \times U(1)'$
gauge symmetry~\cite{anomaly, LL2} due to the anomaly inflow~\cite{Callan, Green}.
 Therefore, the anomaly free condition is that
the sum of the anomaly from the chiral zero modes of the Higgs on the 4-brane at $z=0$, the Standard Model
fermions and exotic particles on the
3-branes at $(y=0, z=0)$ and $(y=\pi R_1/2, z=0)$ for
the gauge group $SU(3)_C\times SU(2)_L\times U(1)_Y \times U(1)'$ should be
zero.

Now, we discuss the Standard Model fermions and exotic particles assignment. 
In order to avoid the proton 
decay problem from the non-renormalizable
operators, we put the Standard Model quarks, $Q_i$, $\bar U_i$ and $\bar D_i$
on the 3-brane at $(y=0, z=0)$,
and put the  Standard Model
leptons $L_i$ and $\bar E_i$ which form the triplet $T_{Li}$, and
right handed neutrinos $\bar \nu_i$ on 
the 3-brane at $(y=\pi R_1/2, z=0)$.  We also introduce a singlet $S$ with
$U(1)'$ charge $s$ on the
3-brane at $(y=\pi R_1/2, z=0)$ so that we can generate the $\mu$ term. 
Moreover, in order to avoid the anomaly, we introduce the
exotic particles.
For simplicity, we introduce  $k_3$ copies of $F_i$ and $\bar F_i$
where $i=1, 2, ..., k_3$,
$k_3'$ copies of $F_i'$ and $\bar F_i'$, $k_2'$ copies of $X_i'$ and $\bar X_i'$,
$k_1$ copies of $Y_i$ and $\bar Y_i$,
$k_1'$ copies of $Y_i'$ and $\bar Y_i'$ on the 3-brane at $(y=0, z=0)$.
And we introduce $k_2$ copies of $T_i$ and $\bar T_i$, $k_0$ copies of $Z_i$ and $\bar Z_i$,
$k_0'$ copies of $Z_i'$ and $\bar Z_i'$ on 
the 3-brane at $(y=\pi R_1/2, z=0)$.

The quantum numbers for the Standard Model
fermions and extra exotic particles under the $SU(3)_C\times SU(2)_L\times U(1)_Y \times
U(1)'$ and $SU(3)_C\times SU(3)\times U(1)'$  gauge symmetry are given in 
Tables 8 and 9, respectively.

Similar to the discussions in the subsection 2.2, the anomaly cancellation conditions are:

\begin{eqnarray}
- 3 + 2 k_3 - k_3' = 0~,~\,
\label{AA1}
\end{eqnarray}
\begin{eqnarray}
 ( 2 k_2 - k_2' + 1 ) ( c_1 + c_2 ) + 3 ( -{{c_1}\over 2} + 3 b' ) = 0 ~,~\,
\label{AA2}
\end{eqnarray}
\begin{eqnarray}
2 - 2 k_1 + k_1' - 4 k_2 + k_2' = 0~,~\,
\label{AA3P}
\end{eqnarray}
\begin{eqnarray}
&& 3 b'^{2} - 6(b'+c_2)^2+3(b'+c_1)^2 + c_2^2 - c_1^2\\ \nonumber
&& +k_3((d_3+2s)^2-d_3^2) + k_3' ((d_3'-s)^2-d_3'^{2})  \\ \nonumber
&& + k_2' (d_2'^{2}-(d_2'-s)^2) +k_1((d_1+2s)^2-d_1^2) + k_1' ((d_1'-s)^2-d_1'^{2})=0~,~\,
\label{AA4}
\end{eqnarray}
\begin{eqnarray}
&&-{9\over 8} c_1^3 - 3 ( -{{c_1}\over 2} + \delta )^3 + 18 b'^{3} - 9 ( b' + c_2 )^3
- 9 ( b' + c_1 )^3  \\ \nonumber
&&+k_0 ( d_0^3 - ( d_0 + 2 s )^3 ) + k_0' (d_0'^3 - ( d_0'- s )^3 ) + k_1
( d_1^3 - ( d_1 + 2 s )^3 )  \\ \nonumber
&&+k_1' ( d_1'^3 - ( d_1' - s )^3 ) + 3 k_2 ( d_2^3 - ( d_2 + 2 s )^3 ) +
2 k_2' ( d_2'^3 - ( d_2' - s )^3 )  \\ \nonumber
&&+3 k_3 ( d_3^3 - ( d_3 + 2 s )^3 ) + 3 k_3' ( d_3'^3 - ( d_3' - s )^3)
+ 2 c_1^3 + 2 c_2^3 + 7 s^3 = 0 ~,~\,
\label{AA5}
\end{eqnarray}
\begin{eqnarray}
(2 k_0 - k_0' + 2 k_2 - k_2') (c_1+c_2) + 3(c_2-\delta) = 0~.~\,
\label{AA6P}
\end{eqnarray}

The simple solutions of Eqs. (\ref{AA1}) and (\ref{AA3P}) for $k_i$ and $k_i'$
are
\begin{eqnarray}
k_1=0~,~k_1'=2~,~ k_2=1~,~ k_2'=0~,~k_3=2~,~k_3'=1~,~  \,
\label{SSIIA}
\end{eqnarray}
\begin{eqnarray}
k_1=0~,~k_1'=1~,~ k_2=1~,~ k_2'=1~,~k_3=2~,~k_3'=1~,~  \,
\label{SSIIB}
\end{eqnarray}
\begin{eqnarray}
k_1=0~,~k_1'=0~,~ k_2=1~,~ k_2'=2~,~k_3=2~,~k_3'=1~.~  \,
\label{SSIIC}
\end{eqnarray}
In particular, for the second solution given by Eq. (\ref{SSIIB}),
$X_i'$, $\bar X_i'$,  $Y_i'$ and $\bar Y_i'$ can form one pair of
the triplets $T'_{i}$ and $\bar T'_{i}$ with quantum number
$(1; 3; d_2')$ and $(1; \bar 3; -(d_2'-s))$ respectively under
the gauge group $SU(3)_C\times SU(3)\times U(1)'$ if $d_2'=d_1'$.
Consequently, we can put this pair of
triplets $T'_{i}$ and $\bar T'_{i}$ on the 3-brane at
$(y=\pi R_1/2, z=0)$ instead of putting 
$X_i'$, $\bar X_i'$,  $Y_i'$ and $\bar Y_i'$ on the  3-brane at
$(y=0, z=0)$.

The $U(1)'$ charge for the right handed neutrinos are
$-(-c_1/2+\delta)$, and we consider the following three scenarios:

\vspace{0.4cm}

(I) $\delta=d_2$ and $ -c_1+2d_2-2s=0$.

\vspace{0.4cm}

(II) $\delta = d_2+s$ and $-c_1+ 2d_2+3s=0$.

\vspace{0.4cm}

(III) $\delta = d_2-s$ and $-c_1 + 2 d_2- 4 s=0$.

\vspace{0.4cm}

It is trivial to find the irrational solutions to the
Eqs. (\ref{AA1}-\ref{AA6P}).
Mathematically speaking, the rational $U(1)'$ charges for all the particles
are equivalent to the integer $U(1)'$ charges for all the particles.
As an example, we present the sample rational solutions to the 
Eqs. (\ref{AA1}-\ref{AA6P})
 in Tables 10, and 11
for above three scenarios where the $k_1$, $k_1'$, $k_2$, $k_2'$,
$k_3$ and $k_3'$ are given in Eqs. (\ref{SSIIA}), and
(\ref{SSIIC}), respectively.
If $k_0=0$ or $k_0'=0$, we do not have the exotic particles
$Z_i$ and $\bar Z_i$ or $Z'_i$ and $\bar Z'_i$. Thus,
$d_0$ or $d'_0$ is not relevant. For simplicity,
we write $d_0=X$ or $d'_0=X$ if $k_0=0$ or $k_0'=0$ in the Tables.

\subsection {Gauge Coupling Unification}
As an example, we discuss the gauge coupling unification for the
scenario II given in Table 10. Since we have one pair of $6$ and ${\bar 6}$
Higgs rather than one pair of $3$ and ${\bar 3}$
Higgs under the $SU(3)$ gauge symmetry on the 4-brane
at $z=0$, the relative running between the $SU(3)_C$ and $SU(3)$ gauge couplings 
are much faster than the previous example in the subsection 3.3. 
Fortunately, the exotic fields help to accelarate considerably the
relative running between the $SU(2)_L$ and $U(1)_Y$ gauge couplings, and 
make the unification possible for TeV scale compactification.
We find that the gauge coupling unification can be
achieved to a good precision (less than $1\%$) if $1/R_1 \simeq 3-4$ TeV with
$1/R_2 \simeq 12/R_1$ and $M_* \simeq 28/R_1$. Notice that
because of the large Casimir operator for the $6$ and ${\bar 6}$
representations ($5/2$), the $SU(3)$ gauge theory is not asymptotically
free above the compactification scale of the fifth dimension
($1/R_1$). The gauge coupling $\alpha$ 
is around $1/22$ at the unification scale $M_*$ which implies
in this scenario $\alpha^\prime_1 \simeq 1/40$ for $U(1)'$ at energy scale
 $1.2$ TeV.

The discussions of gauge coupling unification
for all three scenarios in Tables 10 and 11 are similar. In short,
the gauge coupling unification can be achieved at the $100\sim 200$ TeV
if the compactification scale for the fifth dimension ($1/R_1$)
is about $3\sim 4$ TeV.

\subsection {$SU(3)_C\times SU(2)_L\times U(1)_Y\times U(1)'$
Gauge Symmetry Breaking and Particle Masses}

The discussions for $SU(3)_C\times SU(2)_L\times U(1)_Y\times U(1)'$
gauge symmetry breaking are similar to those in the subsection 3.4 and
Ref.~\cite{JPT}, so,
we do not repeat it here. We would like to emphasize that:
(1) In our scenario, the VEVs for $H_1^0$, $H_2^0$ and $S$ are
at the order of 100 GeV, and the VEVs for $S_i$ are at the order of
1 TeV, then, the $\mu$ problem is solved; (2) We put the
Higgs $\Phi_u$, $\Phi_d$ and $S_i$ on the 4-brane at $z=0$,
and put the singlet $S$ on the 3-brane at $(y=\pi R_1/2, z=0)$. Thus,
we obtain $h\sim 10 \lambda$ and generate the $Z-Z'$ mass hierarchy naturally,
because for instance, in the scenario II in Table 10, $(\pi R_1 M_*)^{1/2} \sim 9.4$.

As an example, we discuss the potential with the $U(1)'$ charges for the Higgs
 and the low energy $U(1)'$ gauge coupling
in the scenario II given in Table 10 in
the last subsection.
With the input
$h=0.75$, $\lambda = 0.075$
$A_h = A_{\lambda} =1.0,
m_{H_d}^2 = m_{H_u}^2= m_{S}^2 = -0.010,
m_{S_1}^2 = m_{S_2}^2=0.031, m_{S_3}^2 = -0.010,
m_{S S_1}^2= m_{S S_2}^2=-0.010, c_1=-15/29,
c_2=-12/29$,  and $s=27/29$, we obtain that
the VEVs for Higgs fields at the  minimum are 
$<H_d> =0.628, <H_u>=0.627,
<S>=0.951, <S_1>=6.45, <S_2>=6.43$, and $<S_3>=6.42$.
The input parameters with dimensions of mass or mass-squared are chosen
in arbitrary units. After finding an acceptable minimum, they are rescaled
so that $\sqrt{v_1^2+v_2^2}\simeq 174$ GeV. After rescaling, the 
VEVs for Higgs fields at the minimum are 
$<H_d> =123.1$ GeV, $<H_u>=122.9$ GeV,
$<S>=186.4$ GeV, $<S_1>=1264.2$ GeV, $<S_2>=1260.7$ GeV, and $<S_3>=1258.9$ GeV.
Thus, we obtain that $\mu= h<S>=139.8$ GeV, $M_Z=91.37$ GeV,
$M_{Z'}=2283$ GeV and $\theta_{Z-Z'}= 0.13\times 10^{-3}$.

First, we discuss the Yukawa couplings for the 
Standard Model quarks and leptons, and the masses of exotic particles
(include $T_i'$ and $\bar T_i'$).
 The 3-brane localized superpotential is
\begin{eqnarray}
S&=&\int d^4x dy dz \delta(y) \delta(z) \int d^2\theta
\left(
  y_{uij}^5 Q_i H_u {\bar U_j}+  y_{dij}^5 Q_i H_d {\bar D_j}
\right.\nonumber\\&&\left. 
+ h_{Fi}^5 S_3 F_i {\bar F_i} +  h^{\prime5}_{Fi} S_1 F_i' {\bar F_i'}
+{\lambda}^{\prime5}_{Fi} S_2 F_i' {\bar F_i'}+  h^{\prime 5}_{Xi} S_1 X_i' {\bar X_i'}
\right.\nonumber\\&&\left.
+{\lambda}^{\prime5}_{Xi} S_2 X_i' {\bar X_i'}
+ h_{Yi}^5 S_3 Y_i {\bar Y_i} +  h^{\prime5}_{Yi} S_1 Y_i' {\bar Y_i'}
+{\lambda}^{\prime5}_{Yi} S_2 Y_i' {\bar Y_i'}+ H. C. \right)
\nonumber\\&&
+ \int d^4x dy dz \delta(y-\pi R_1) \delta(z) \int d^2\theta
\left( y_{eij}^5 T_{Li} \Phi_d T_{Li}
\right.\nonumber\\&&\left.
+ h_{Ti}^5 S_3 T_i {\bar T_i} + h^{\prime5}_{Ti} S_1 T_i' {\bar T_i'}
+{\lambda}^{\prime5}_{Ti} S_2 T_i' {\bar T_i'}
\right.\nonumber\\&&\left.
+ h_{Zi}^5 S_3 Z_i {\bar Z_i} +  h^{\prime5}_{Zi} S_1 Z_i' {\bar Z_i'}
+{\lambda}^{\prime5}_{Zi} S_2 Z_i' {\bar Z_i'}
 + H. C. \right)
~.~\,
\end{eqnarray}  
In the 4-dimensional effective theory, we have
the following superpotential
\begin{eqnarray}
W&=& y_{uij} Q_i H_u {\bar U_j} +  y_{dij} Q_i H_d {\bar D_j}
+ y_{eij} L_i H_d {\bar E_j} 
\nonumber\\&&
+ h_{Fi} S_3 F_i {\bar F_i} +  h^{\prime}_{Fi} S_1 F_i' {\bar F_i'}
+{\lambda}^{\prime}_{Fi} S_2 F_i' {\bar F_i'}
\nonumber\\&&
+ h_{Ti} S_3 T_i {\bar T_i} +  h^{\prime}_{Xi} S_1 X_i' {\bar X_i'}
+{\lambda}^{\prime}_{Xi} S_2 X_i' {\bar X_i'}
\nonumber\\&&
+ h_{Yi} S_3 Y_i {\bar Y_i} +  h^{\prime}_{Yi} S_1 Y_i' {\bar Y_i'}
+{\lambda}^{\prime}_{Yi} S_2 Y_i' {\bar Y_i'}
\nonumber\\&&
+ h_{Zi} S_3 Z_i {\bar Z_i} +  h^{\prime}_{Zi} S_1 Z_i' {\bar Z_i'}
+{\lambda}^{\prime}_{Zi} S_2 Z_i' {\bar Z_i'}
\nonumber\\&&
 +  h^{\prime}_{Ti} S_1 T_i' {\bar T_i'}
+{\lambda}^{\prime}_{Ti} S_2 T_i' {\bar T_i'}
~.~\,
\end{eqnarray}  
Therefore, all the exotic particles can obtain masses around 1 TeV
after the gauge symmetry breaking because the VEVs for $S_i$ are at the order of 1 TeV.

Second, we would like to discuss the neutrino masses.
Because the Higgs sextets $\Phi_u$ and $\Phi_d$ can not give the neutrino
Dirac masses, we assume that $T_i$ obtain a small VEV
 compare to the VEVs of $H_u$ and $H_d$\footnote{$T_i'$
might also obtain the VEV in general. If this is case, we assume that
the VEV of $T_i'$ is much smaller than that of $T_i$, similar to 
the large ${\rm tan}\beta$ scenario for $H_u$ and $H_d$ in the Minimal Supersymmetric
Standard Model.}. 
Note that $T_i$ can not give masses
to the leptons and quarks. Because in our models, we
just introduce one pair of $T_i$ and $\bar T_i$, we write
$T$ for $T_i$ to simplify the notations.
Similar to the model in Section 3, the dimension-5 operators  
$y_{\nu ij} T_{Li} T_{Li} T T/M_*$ are forbidden by
$U(1)'$ gauge symmetry.
To be explicit, we will use $<T^0> \sim 10$ GeV as an example
to discuss the neutrino masses for three scenarios with different $\delta$.

(I) $\delta=d_2$ and $ -c_1+2d_2-2s=0$.
The neutrinos can obtain masses via the localized superpotential
\begin{eqnarray}
S&=&\int d^4x dy dz \delta(y-\pi R_1) \delta(z) \left[\int d^2\theta
h_{rij}^5 S_3 {\bar \nu_i} {\bar \nu_j} + y_{\nu ij}^5 T T_{Li} {\bar \nu_j}  + H. C. \right]
~,~\,
\end{eqnarray}
or in 4-dimensional effective theory, the superpotential is
\begin{eqnarray}
W&=& h_{rij} S_3 {\bar \nu_i} {\bar \nu_j} + y_{\nu ij} T T_{Li} {\bar \nu_j} 
~.~\,
\end{eqnarray}
Because in our model, $<S_3>$ is at the order of 1 TeV, the 
right-handed neutrino masses are about 1 TeV if $h_{rij}$ is
at the order of 1. Therefore, in order to have the realistic
active neutrino masses, we obtain that the neutrino Dirac masses should be
at the order of $10^{-3}$ GeV, {\it i.e.}, the Yukawa couplings
($y_{\nu ij}$)  should
be at the order of $10^{-4}\sim 10^{-5}$ or less. 

(II) $\delta = d_2+s$ and $-c_1+ 2d_2+3s=0$.
The neutrinos can obtain masses via the superpotential in the 4-dimensional
effective theory
\begin{eqnarray}
W&=& h_{rij} S_1 {\bar \nu_i} {\bar \nu_j}
+  h'_{rij} S_2 {\bar \nu_i} {\bar \nu_j}
 + y_{\nu ij} {S\over M_*} T T_{Li} {\bar \nu_j} 
~.~\,
\end{eqnarray}
In our model, $<S_1>$ and $<S_2>$ are at the order of 1 TeV, then, the 
right-handed neutrino masses are about 2 TeV if $h_{rij}$ and $h'_{rij}$ is
at the order of 1.
In order to have the realistic
active neutrino masses, we obtain that the couplings $y_{\nu ij}$ for the non-renormalizable
terms should be at the order of 0.2 or smaller than 0.2 since $M_* \sim 10^5$ GeV.

(III) $\delta = d_2-s$ and $-c_1 + 2 d_2- 4 s=0$.
The neutrinos can obtain masses via the superpotential in the 4-dimensional
effective theory
\begin{eqnarray}
W&=& h_{rij} S_3 {\bar \nu_i} {\bar \nu_j} 
 + {{\sqrt 2} \over {(\pi R_1 M_*)^{1/2}}}
( y_{\nu ij}  {S_1\over M_*}
+ y'_{\nu ij} {S_2\over M_*}) T T_{Li} {\bar \nu_j}
~.~\,
\end{eqnarray}
For the scenario II in Table 10 in the last subsection,
$(\pi R_1 M_*/2)^{1/2}\sim 6.63$.
The right-handed neutrino masses are about 2 TeV if $h_{rij}$ is
about 2. And  in order to have the realistic
active neutrino masses, we obtain that
the couplings $y_{\nu ij}$ and $y'_{\nu ij}$  for the non-renormalizable
terms should be at the order of or smaller than 0.1.
If there is cancellation in $(y_{\nu ij}  <S_1> + y'_{\nu ij} <S_2>)$,
the couplings $y_{\nu ij}$ and $y'_{\nu ij}$ can be at the order of 1.

In short, the very small realistic neutrino masses can be generated naturally
in scenarios (II) and (III).

\section{Discussion and Conclusion}
There exist some variants of our models. For instance, for the model
in Section 4, we can put one Higgs sextet $\Psi_d$, two singlets
$\Psi_{S_1}$ and $\Psi_{S_2}$ on the 4-brane at $z=0$, and put 
one Higgs doublet $H_u$, two singlets $S$ and $S_1$ on the 3-brane
at $(y=0, z=0)$. The anomaly free conditions are the same, and the
discussions for the gauge coupling unification, gauge symmetry breaking, $\mu$ problem,
$Z-Z'$ mass hierarchy and neutrino masses, etc., are similar.

In addition, we can discuss the models with the general exotic particles.
To be explicit, we can add the exotic particles, which transfrom as
(3;, 2; 1/6), (1; 2; -1/2), ($\bar 3$; 1; -2/3), ($\bar 3$; 1; 1/3),
(1; 1;, 1), (1; 3; 0) under the $SU(3)_C\times SU(2)_L \times U(1)_Y$ 
gauge symmetry, and
their complex conjugation fields (mirror partners) under the Standard
Model gauge symmetry~\cite{Dobrescu, Erler}. 
 Similarly, one can calculate the anomaly free
conditions and obtain the anomaly free models. In the
models with general exotic
particles, it is relatively easy to obtain the anomaly free models
where the particles have rational $U(1)'$ charges because one has more
freedoms~\cite{Dobrescu, Erler, EMa}. This kind of generalizations
is similar and straightforward, so, we do not consider it here.

In this paper, 
 we consider the low energy 6-dimensional $N=2$ supersymmetric $SU(6)$ gauge 
unification theory on the space-time $M^4\times T^2/(Z_2)^3$.
First, we discuss the orbifold gauge symmetry breaking, which breaks
the $SU(6)$ down to the $SU(3)_C \times SU(2)_L \times U(1)_Y \times U(1)'$
gauge symmetry for the zero modes.  Then, we
present the parity assignment and masses for the bulk gauge fields and
Higgs fields on the 4-brane at $z=0$, and the number of the 4-dimensional
supersymmetry and gauge symmetry on the 3-branes at the fixed points and
on the 4-branes on the fixed lines. Second, we discuss the
anomaly cancellation. In order to cancel the anomalies involving
at least one $U(1)'$, we add extra exotic particles which
are vector-like under the Standard Model gauge symmetry. And we study
the anomaly free conditions and give
some anomaly free models. Third, similar to the discussions
in Ref.~\cite{JPT}, we discuss the $SU(3)_C\times SU(2)_L\times U(1)_Y \times U(1)'$
gauge symmetry breaking and solve the $\mu$ problem.
The $Z-Z'$ mass hierarchy can be generated naturally because
we can have $h \sim 10 \lambda$ if we
put the Higgs triplets $\Phi_u$, $\Phi_d$ and singlet $S_i$ on the 4-brane at $z=0$
and put the singlet $S$ on the 3-brane at $(y=\pi R_1, z=0)$.
 Fourth, we discuss the gauge coupling unification, which shows that
the gauge couplings can be unified at $100\sim 200$ TeV if the compactification
scale for the fifth dimension is $3\sim 4$ TeV. The proton decay
problem can be avoided by putting the quarks and leptons/neutrinos on the
different 3-branes.
Fifth, we discuss the masses for the extra exotic particles,
which can be at the order of 1 TeV after the gauge symmetry breaking, and the masses
for the extra zero modes of the chiral superfields from the vector
multiplets and Higgs hypermultiplets.
In particular, we discuss the neutrino masses in detail.
We forbid the dimension-5 operators $y_{\nu ij} L_i L_j H_u H_u/M_*$
by $U(1)'$ gauge symmetry, and the
correct active neutrino masses can be obtained via the non-renormalizable
terms.

Next, we consider the low energy 6-dimensional $N=2$ supersymmetric $SU(6)$ gauge 
unification theory on the space-time $M^4\times T^2/(Z_2)^4$.
First, we discuss the orbifold gauge symmetry breaking, then, we
present the parity assignment and masses for the bulk gauge fields and
Higgs fields on the 4-brane at $z=0$, and the number of the 4-dimensional
supersymmetry and gauge symmetry on the 3-branes at the fixed points and
on the 4-branes on the fixed lines. We would like to point out that,
in order to avoid the proton decay
problem, we  put the quarks on the 3-brane at $(y=0, z=0)$
 and leptons/neutrinos on the 3-brane at  $(y=\pi R_1/2, z=0)$,
which preserve the $SU(3)_C\times SU(2)_L\times U(1)_Y \times U(1)'$ and
$SU(3)_C\times SU(3)\times U(1)'$ gauge symmetry, respectively.
And in order to generate the correct lepton masses, we add one 
pair of Higgs sextets on the 4-brane at $z=0$, instead of one pair of triplets.
 Second, we discuss the
anomaly cancellation by adding the extra exotic particles which
are vector-like under the Standard Model gauge symmetry. We also present
 the anomaly free conditions and
some anomaly free models. Third, simlar to 
the discussions in Ref.~\cite{JPT},  the $SU(3)_C\times SU(2)_L\times U(1)_Y \times U(1)'$
gauge symmetry can be broken and the $\mu$
problem can be solved. The $Z-Z'$ mass hierarchy can be generated naturally because
we can have $h \sim 10 \lambda$ if we
put the Higgs sextets $\Phi_u$, $\Phi_d$ and singlet $S_i$ on the 4-brane at $z=0$
and put the singlet $S$ on the 3-brane at $(y=\pi R_1/2, z=0)$.
 Fourth, we  show that
the gauge couplings can be achieved at $100\sim 200$ TeV if the compactification
scale for the fifth dimension is $3\sim 4$ TeV. 
Fifth, we discuss the masses for the extra exotic particles,
which can be at the order of 1 TeV after the gauge symmetry breaking.
Especially, in order to have the correct neutrino masses, we 
give the small VEVs to the triplet exotic particle
$T$  on the 3-brane at $(y=\pi R_1/2, z=0)$.
We forbid the dimension-5 operators $y_{\nu ij} T_{Li} T_{Li} T T/M_*$
by $U(1)'$ gauge symmetry,
  and  the realistic active neutrino masses can be
 generated via the non-renormalizable terms.

\section*{Acknowledgments}
The research of T. Li was supported  by the National Science Foundation under
 Grant No.~PHY-0070928,
and the research of J. Jiang was supported  by the U.S.~Department of Energy under Grant 
 No.~W-31-109-ENG-38.

\newpage

\renewcommand{\arraystretch}{1.4}
\begin{table}[t]
\caption{Parity assignment and masses ($n\ge 0, m \ge 0$) for the vector multiplet 
and Higgs fields in 
 the $SU(6)$ model on $M^4\times T^2/(Z_2)^3$.
\label{tab:SUV1}}
\vspace{0.4cm}
\begin{center}
\begin{tabular}{|c|c|c|}
\hline        
$(P^y, P^{z}, P^{z'})$ & {\rm Fields} & {\rm Mass}\\ 
\hline
$(+, +, +)$ &  $V^{a, b}_{\mu}$, $\Sigma_5^{\hat{a},b}$ & $\sqrt {n^2/R_1^2+ (2m)^2/R_2^2}$ \\
\hline
$(+, +, -)$ &  $\Sigma_6^{a, \hat{b}}$, $\Phi^{\hat{a},\hat{b}}$ & $\sqrt {n^2/R_1^2+ (2m+1)^2/R_2^2}$ \\
\hline
$(+, -, +)$ &  $V^{a, \hat{b}}_{\mu}$, $\Sigma_5^{\hat{a},\hat{b}}$ & $\sqrt {n^2/R_1^2+ (2m+1)^2/R_2^2}$ \\
\hline
$(+, -, -)$ &  $\Sigma_6^{a, b}$, $\Phi^{\hat{a},b}$ & $\sqrt {n^2/R_1^2+ (2m+2)^2/R_2^2}$ \\
\hline
$(-, +, +)$ &  $V^{\hat{a}, b}_{\mu}$, $\Sigma_5^{a,b}$ & $\sqrt {(n+1)^2/R_1^2+ (2m)^2/R_2^2}$ \\
\hline
$(-, +, -)$ &  $\Sigma_6^{\hat{a}, \hat{b}}$, $\Phi^{a,\hat{b}}$ & $\sqrt {(n+1)^2/R_1^2+ (2m+1)^2/R_2^2}$ \\
\hline
$(-, -, +)$ &  $V^{\hat{a}, \hat{b}}_{\mu}$, $\Sigma_5^{a,\hat{b}}$ & $\sqrt {(n+1)^2/R_1^2+ (2m+1)^2/R_2^2}$\\
\hline
$(-, -, -)$ &  $\Sigma_6^{\hat{a}, b}$, $\Phi^{a,b}$ & $\sqrt {(n+1)^2/R_1^2+ (2m+2)^2/R_2^2}$ \\
\hline
$P^y=+$ & $H_u$, $S_u^c$, $H_d$, $S_d^c$, $S_i$ & $n/R_1$\\
\hline
$P^y=-$ & $H_u^c$, $S_u$, $H_d^c$, $S_d$, $S_i^c$ & $(n+1)/R_1$\\
\hline
\end{tabular}
\end{center}
\end{table}

\renewcommand{\arraystretch}{1.4}
\begin{table}[t]
\caption{For the $SU(6)$ model on $M^4\times T^2/(Z_2)^3$, 
the gauge superfields, the
number of 4-dimensional supersymmetry and gauge symmetry on the 3-brane, which
is located at the fixed point $(y=0, z=0),$ $(y=0, z=\pi R_2/2),$ $(y=\pi R_1, z=0)$, or 
$(y=\pi R_1, z=\pi R_2/2)$, and on the 4-brane which is located on the fixed line
$y=0$, $z=0$, $y=\pi R_1$, or $z=\pi R_2/2$.
\label{tab:SUV2}}
\vspace{0.4cm}
\begin{center}
\begin{tabular}{|c|c|c|c|}
\hline        
${\rm Brane~ Position}$ & ${\rm Fields}$ & ${\rm SUSY}$ & ${\rm Gauge ~Symmetry}$\\ 
\hline
$(0, 0) $ &  $V^{a,b}_{\mu}$, $\Sigma_5^{\hat a, b}$,
 $\Sigma_6^{a, \hat b}$, $\Phi^{\hat a, \hat b}$
  & N=1 & $SU(3)\times SU(2)\times U(1) \times U(1)$ \\
\hline
$(0, \pi R_2/2)$ & $V^{a,B}_{\mu}$, $\Sigma_5^{\hat a, B}$  & N=1 &
 $SU(4)\times SU(2)\times U(1)$ \\
\hline
$(\pi R_1, 0) $ & $V^{a,b}_{\mu}$, $\Sigma_5^{\hat a, b}$,
 $\Sigma_6^{a, \hat b}$, $\Phi^{\hat a, \hat b}$
  & N=1 & $SU(3)\times SU(2)\times U(1) \times U(1)$ \\
\hline
$(\pi R_1, \pi R_2/2) $ & $V^{a,B}_{\mu}$, $\Sigma_5^{\hat a, B}$  & N=1 &
 $SU(4)\times SU(2)\times U(1) $ \\
\hline
$y=0$ &  $V^{a,B}_{\mu}$, $\Sigma_5^{\hat a, B}$, $\Sigma_6^{a, B} $, $\Phi^{\hat a, B}$
  & N=2 & $SU(4)\times SU(2)\times U(1)$ \\
\hline
$z=0$ &  $V^{A,b}_{\mu}$, $\Sigma_5^{A,b}$, $\Sigma_6^{A, \hat b} $, $\Phi^{A, \hat b}$
  & N=2 & $SU(3)\times SU(3)\times U(1)$ \\
\hline
$y=\pi R_1 $ & $V^{a,B}_{\mu}$, $\Sigma_5^{\hat a, B}$, $\Sigma_6^{a, B} $, $\Phi^{\hat a, B}$
  & N=2 & $SU(4)\times SU(2)\times U(1)$ \\
\hline
$z=\pi R_2/2$ & $V^{A,B}_{\mu}$, $\Sigma_5^{A,B}$  & N=2 & $SU(6)$ \\
\hline
\end{tabular}
\end{center}
\end{table}

\renewcommand{\arraystretch}{1.4}
\begin{table}[t]
\caption{Quantum numbers for the Standard Moderl
fermions ($Q_i$, $\bar U_i$, $\bar D_i$, $L_i$, $\bar \nu_i$, $\bar E_i$)
and extra exotic particles ($F_i$, $\bar F_i$, $F_i'$, $\bar F_i'$,
$X_i$, $\bar X_i$, $X_i'$, $\bar X_i'$,
$Y_i$, $\bar Y_i$, $Y_i'$, $\bar Y_i'$,
$Z_i$, $\bar Z_i$, $Z_i'$, $\bar Z_i'$)
under the $SU(3)_C\times SU(2)_L\times U(1)_Y \times
U(1)'$ gauge symmetry in the $SU(6)$ model on $M^4\times T^2/(Z_2)^3$.
\label{tab:SUV3}}
\vspace{0.4cm}
\begin{center}
\begin{tabular}{|c|c|c|c|}
\hline        
Particles &  Quantum Numbers & Particles &  Quantum Numbers \\ 
\hline
$L_i$ & (1; 2; $-1/2$; $a'$) &
$Q_i$ & (3; 2; 1/6; $b'$) \\
\hline
$\bar \nu_i$ & (1; 1; 0; $-(a'+\delta))$ & $\bar U_i$ & ($\bar 3$; 1; $-2/3$;  $-(b'+c_2)$)\\
\hline
$\bar E_i$ & (1; 1; 1; $-(a'+c_1)$) & $\bar D_i$ & ($\bar 3$; 1; 1/3;  $-(b'+c_1)$)\\
\hline
\hline
$S$ & (1; 1; 0; $s$) & & \\
\hline
$F_i$ & (3; 1; $-1/3$; $d_3$) & $\bar F_i$ & ($\bar 3$; 1; 1/3;  $-(d_3+2s)$)\\
\hline
$F_i'$ & (3; 1; $-1/3$; $d_3'$) & $\bar F_i'$ & ($\bar 3$; 1; 1/3;  $-(d_3'-s)$)\\
\hline
$X_i$ & (1; 2; 1/2; $d_2$) & $\bar X_i$ & (1; 2; $-1/2$;  $-(d_2+2s)$)\\
\hline
$X_i'$ & (1; 2; 1/2; $d_2'$) & $\bar X_i'$ & (1; 2; $-1/2$;  $-(d_2'-s)$)\\
\hline
$Y_i$ & (1; 1; $-1$; $d_1$) & $\bar Y_i$ & (1; 1; 1;  $-(d_1+2s)$)\\
\hline
$Y_i'$ & (1; 1; $-1$; $d_1'$) & $\bar Y_i'$ & (1; 1; 1;  $-(d_1'-s)$)\\
\hline
$Z_i$ & (1; 1; 0; $d_0$) & $\bar Z_i$ & (1; 1; 0;  $-(d_0+2s)$)\\
\hline
$Z_i'$ & (1; 1; 0; $d_0'$) & $\bar Z_i'$ & (1; 1; 0;  $-(d_0'-s)$)\\
\hline
\end{tabular}
\end{center}
\end{table}

\begin{table}[ht]
\caption{Sample: rational $U(1)'$ charges of all the particles for the solution
with  $k_1=2$, $k_1'=0$, $k_2=0$, $k_2'=1$, $k_3=2$ and $k_3'=1$
 in the $SU(6)$ model on $M^4\times T^2/(Z_2)^3$.
\label{tab:suv5}}
\vspace{0.4cm}
\begin{center}
\begin{tabular}{|c|c|c|c|c|c|c|c|c|c|c|c|c|}
\hline        
 Scenario & $k_0$ & $k_0'$ & $d_0$ & $d_0'$ & $d_1$ & $d_3$ & $d_2'$ & $d_3'$ &
 $a'$ & $b'$ & $c_1$ & $c_2$  \\
\hline
 I & 1 & 1 & -1 & 1 & 1 & 1 & -1 & -1 & -$\frac{15}{8}$ &
 $\frac{5}{8}$ & $\frac{7}{8}$ & $\frac{1}{2}$ \\
 II & 0 & 2 & X & -1 & 1 & 1 & -1 & -1 & $\frac{2}{3}$ &
 -$\frac{2}{9}$ & $\frac{1}{6}$ & $\frac{5}{6}$ \\
 III & 0 & 8 & X & 1 & -1 & -1 & -1 & 1 & -1 & $\frac{1}{3}$ & 1 & -3 \\
 IV & 2 & 0 & 1 & X & 1 & 1 & -1 & 1 & $\frac{7}{4}$ & -$\frac{7}{12}$
 & 1 & -$\frac{5}{4}$ \\
 V & 4 & 1 & 1 & 1 & -1 & 1 & -1 & -1 & -$\frac{11}{8}$ &
 $\frac{11}{24}$ & $\frac{1}{8}$ & $\frac{1}{4}$ \\
\hline
\end{tabular}
\end{center}
\end{table}

\begin{table}[ht]
\caption{Sample: rational $U(1)'$ charges of all the particles for the solution
with   $k_1=1$, $k_1'=1$, $k_2=1$, $k_2'=0$, $k_3=2$ and $k_3'=1$
 in the $SU(6)$ model on $M^4\times T^2/(Z_2)^3$.
\label{tab:suv6}}
\vspace{0.4cm}
\begin{center}
\begin{tabular}{|c|c|c|c|c|c|c|c|c|c|c|c|c|c|}
\hline        
 Scenario & $k_0$ & $k_0'$ & $d_0$ & $d_0'$ & $d_1$ & $d_2$ & $d_3$ & $d_1'$ & $d_3'$ &
 $a'$ & $b'$ & $c_1$ & $c_2$  \\
\hline
 I & 1 & 4 & -1 & -1 & -1 & -1 & 1 & 1 & 1 & 3 & -$\frac{2}{3}$ & 1 & -2 \\
 II & 0 & 5 & X & 1 & -1 & 1 & -1 & -1 & 1 & -$\frac{13}{9}$ &
 $\frac{61}{81}$ & -$\frac{28}{27}$ & $\frac{2}{9}$ \\
 III & 0 & 11 & X & -1 & -1 & 1 & -1 & -1 & 1 & -$\frac{2}{3}$ &
 $\frac{13}{18}$ & $\frac{5}{6}$ & -$\frac{7}{3}$ \\
 IV & 1 & 1 & -1 & -1 & 1 & 1 & -1 & 1 & -1 & -1 & $\frac{3}{8}$ &
 -$\frac{11}{8}$ & $\frac{5}{4}$ \\
 V & 2 & 0 & -1 & X & 1 & 1 & -1 & -1 & 1 & -$\frac{1}{3}$ &
 $\frac{2}{9}$ & -$\frac{5}{3}$ & $\frac{4}{3}$ \\
\hline
\end{tabular}
\end{center}
\end{table}

\renewcommand{\arraystretch}{1.4}
\begin{table}[t]
\caption{Parity assignment and masses ($n\ge 0, m \ge 0$) for the
gauge and Higgs fields in the $SU(6)$ model on $M^4\times T^2/(Z_2)^4$.
\label{tab:SUV9}}
\vspace{0.4cm}
\begin{center}
\begin{tabular}{|c|c|c|}
\hline        
$(P^y, P^{y'}, P^z, P^{z'})$ & field & mass\\ 
\hline
$(+, +, +, +)$ &  $V^{a, b}_{\mu}$ & $\sqrt {(2n)^2/R_1^2+ (2m)^2/R_2^2}$ \\
\hline
$(+, +, -, +)$ &  $V^{a, {\hat b}}_{\mu}$ & $\sqrt {(2n)^2/R_1^2+ (2m+1)^2/R_2^2}$ \\
\hline
$(-, +, +, +)$ &  $V^{{\hat a}, b}_{\mu}$ & $\sqrt {(2n+1)^2/R_1^2+ (2m)^2/R_2^2}$ \\
\hline
$(-,+, -, +)$ &  $V^{\hat{a}, \hat{b}}_{\mu}$ & $\sqrt {(2n+1)^2/R_1^2+(2m+1)^2/R_2^2}$ \\
\hline
$(-, -, +, +)$ &  $\Sigma_5^{a, b}$ & $\sqrt {(2n+2)^2/R_1^2+ (2m)^2/R_2^2}$ \\
\hline
$(-, -, -, +)$ &  $\Sigma_5^{a, {\hat b}}$ & $\sqrt {(2n+2)^2/R_1^2+ (2m+1)^2/R_2^2}$ \\
\hline
$(+, -, +, +)$ &  $\Sigma_5^{{\hat a}, b}$ & $\sqrt {(2n+1)^2/R_1^2+ (2m)^2/R_2^2}$ \\
\hline
$(+, -, -, +)$ &  $\Sigma_5^{\hat{a}, \hat{b}}$ & $\sqrt {(2n+1)^2/R_1^2+ (2m+1)^2/R_2^2}$ \\
\hline
$(+, +, -, -)$ &  $\Sigma_6^{a, b}$ & $\sqrt {(2n)^2/R_1^2+ (2m+2)^2/R_2^2}$\\
\hline
$(+, +, +, -)$ &  $\Sigma_6^{a, {\hat b}}$ & $\sqrt {(2n)^2/R_1^2+ (2m+1)^2/R_2^2}$\\
\hline
$(-, +, -, -)$ &  $\Sigma_6^{{\hat a}, b}$ & $\sqrt {(2n+1)^2/R_1^2+ (2m+2)^2/R_2^2}$\\
\hline
$(-, +, +,  -)$ &  $\Sigma_6^{\hat{a}, {\hat b}}$ & $\sqrt {(2n+1)^2/R_1^2+ (2m+1)^2/R_2^2}$ \\
\hline
$(-, -, -, -)$ &  $\Phi^{a, b}$ & $\sqrt {(2n+2)^2/R_1^2+ (2m+2)^2/R_2^2}$\\
\hline
$(-, -, +, -)$ &  $\Phi^{a, {\hat b}}$ & $\sqrt {(2n+2)^2/R_1^2+ (2m+1)^2/R_2^2}$\\
\hline
$(+, -, -, -)$ &  $\Phi^{{\hat a}, b}$ & $\sqrt {(2n+1)^2/R_1^2+ (2m+2)^2/R_2^2}$\\
\hline
$(+, -, +, -)$ &  $\Phi^{\hat{a}, {\hat b}}$ & $\sqrt {(2n+1)^2/R_1^2+(2m+1)^2/R_2^2}$\\
\hline
$(p^y=+, P^{y'}=+)$ & $H_u$, $H_d$, $S_i$ & $2n/R_1$ \\
\hline
$(p^y=-, P^{y'}=+)$ & $\eta_u$, $\eta_d$ & $(2n+1)/R_1$ \\
\hline
$(p^y=+, P^{y'}=-)$ & $\eta_u^c$, $\eta_d^c$ & $(2n+1)/R_1$ \\
\hline
$(p^y=-, P^{y'}=-)$ & $H_u^c$, $H_d^c$, $S_i^c$ & $(2n+2)/R_1$ \\
\hline
\end{tabular}
\end{center}
\end{table}

\renewcommand{\arraystretch}{1.4}
\begin{table}[t]
\caption{For the $SU(6)$ model on $M^4\times T^2/(Z_2)^4$, the gauge
superfields, the number of 4-dimensional supersymmetry and gauge symmetry on the 3-brane, which
is located at the fixed point $(y=0, z=0),$ $(y=0, z=\pi R_2/2),$ $(y=\pi R_1/2, z=0)$, or 
$(y=\pi R_1/2, z=\pi R_2/2)$, and on the 4-brane which is located on the fixed line
$y=0$, $z=0$, $y=\pi R_1/2$, or $z=\pi R_2/2$.
\label{tab:SUV10}}
\vspace{0.4cm}
\begin{center}
\begin{tabular}{|c|c|c|c|}
\hline        
Brane Position & fields & SUSY & Gauge Symmetry\\ 
\hline
$(0, 0) $  & $V^{a,b}_{\mu}$, $\Sigma_5^{\hat a, b}$,
 $\Sigma_6^{a, \hat b}$, $\Phi^{\hat a, \hat b}$
  & N=1 & $SU(3)\times SU(2)\times U(1) \times U(1)$ \\
\hline
$(0, \pi R_2/2)$  & $V^{a,B}_{\mu}$, $\Sigma_5^{\hat a, B}$  & N=1 & $SU(4)\times SU(2) \times U(1) $ \\
\hline
$(\pi R_1/2, 0) $ & $V^{A,b}_{\mu}$, $\Sigma_6^{A, \hat b}$  & N=1 & $SU(3)\times SU(3)\times U(1)$ \\
\hline
$(\pi R_1/2, \pi R_2/2) $ &  $V^{A,B}_{\mu}$ & $N=1$ & $SU(6)$ \\
\hline
$y=0$ & $V^{a,B}_{\mu}$, $\Sigma_5^{\hat a, B}$, $\Sigma_6^{a, B} $, $\Phi^{\hat a, B}$
  & N=2 & $SU(4)\times SU(2) \times U(1)$ \\
\hline
$z= 0 $  & $V^{A,b}_{\mu}$, $\Sigma_5^{A,b}$, $\Sigma_6^{A, \hat b} $, $\Phi^{A, \hat b}$ & N=2 &
 $SU(3)\times SU(3)\times U(1)$ \\
\hline
$y=\pi R_1/2 $ &  $V^{A,B}_{\mu}$, $\Sigma_6^{A,B}$  & N=2 & $SU(6)$ \\
\hline
$z=\pi R_2/2 $  & $V^{A,B}_{\mu}$, $\Sigma_5^{A,B}$  & N=2 & $SU(6)$ \\
\hline
\end{tabular}
\end{center}
\end{table}

\renewcommand{\arraystretch}{1.4}
\begin{table}[t]
\caption{Quantum numbers for the Standard Moderl
quarks ($Q_i$, $\bar U_i$, $\bar D_i$)
and extra exotic particles ($F_i$, $\bar F_i$, $F_i'$, $\bar F_i'$,
 $X_i'$, $\bar X_i'$, $Y_i$, $\bar Y_i$, $Y_i'$ and $\bar Y_i'$)
under the $SU(3)_C\times SU(2)_L\times U(1)_Y \times
U(1)'$ gauge symmetry in the $SU(6)$ model on $M^4\times T^2/(Z_2)^4$.
 These particles are on the 3-brane at $(y=0, z=0)$.
\label{tab:SUV11}}
\vspace{0.4cm}
\begin{center}
\begin{tabular}{|c|c|c|c|}
\hline        
Particles & Quantum Numbers & Particles & Quantum Numbers \\ 
\hline
$Q_i$ & (3; 2; 1/6; $b'$)  & & \\
\hline
$\bar U_i$ & ($\bar 3$; 1; $-2/3$;  $-(b'+c_2)$) & $\bar D_i$ & ($\bar 3$; 1; 1/3;  $-(b'+c_1)$)\\
\hline
\hline
$F_i$ & (3; 1; $-1/3$; $d_3$) & $\bar F_i$ & ($\bar 3$; 1; 1/3;  $-(d_3+2s)$)\\
\hline
$F_i'$ & (3; 1; $-1/3$; $d_3'$) & $\bar F_i'$ & ($\bar 3$; 1; 1/3;  $-(d_3'-s)$)\\
\hline
$X_i'$ & (1; 2; 1/2; $d_2'$) & $\bar X_i'$ & (1; 2; $-1/2$;  $-(d_2'-s)$)\\
\hline
$Y_i$ & (1; 1; $-1$; $d_1$) & $\bar Y_i$ & (1; 1; 1;  $-(d_1+2s)$)\\
\hline
$Y_i'$ & (1; 1; $-1$; $d_1'$) & $\bar Y_i'$ & (1; 1; 1;  $-(d_1'-s)$)\\
\hline
\end{tabular}
\end{center}
\end{table}

\renewcommand{\arraystretch}{1.4}
\begin{table}[t]
\caption{Quantum numbers for the Standard Moderl
leptons $T_{Li}$ and right handed neutrinos $\bar \nu_i$,
and extra exotic particles $T_i$, $\bar T_i$,
$Z_i$, $\bar Z_i$, $Z_i'$, $\bar Z_i'$
under the $SU(3)_C\times SU(3) \times
U(1)'$ gauge symmetry in the $SU(6)$ model on $M^4\times T^2/(Z_2)^4$.
  These particles are on the 3-brane at $(y=\pi R_1/2, z=0)$.
\label{tab:SUV12}}
\vspace{0.4cm}
\begin{center}
\begin{tabular}{|c|c|c|c|}
\hline        
Particles & Quantum Numbers & Particles  & Quantum Numbers \\ 
\hline
$T_{Li}$ & (1; $\bar 3$; $-c_1/2$) &
$\bar \nu_i$ & (1; 1; $-(-c_1/2+\delta))$ \\
\hline
\hline
$S$ & (1; 1; 0; s) & & \\
\hline
$T_i$ & (1; 3; $d_2$) & $\bar T_i$ & (1; $\bar 3$;  $-(d_2+2s)$)\\
\hline
$Z_i$ & (1; 1; 0; $d_0$) & $\bar Z_i$ & (1; 1; 0;  $-(d_0+2s)$)\\
\hline
$Z_i'$ & (1; 1; 0; $d_0'$) & $\bar Z_i'$ & (1; 1; 0;  $-(d_0'-s)$)\\
\hline
\end{tabular}
\end{center}
\end{table}

\renewcommand{\arraystretch}{1.4}
\begin{table}[ht]
\caption{Sample: rational $U(1)'$ charges of all the particles for the solution
with  $k_1=0$, $k_1'=2$, $k_2=1$, $k_2'=0$, $k_3=2$ and $k_3'=1$
 in the $SU(6)$ model on $M^4\times T^2/(Z_2)^4$.
\label{tab:suv13}}
\vspace{0.4cm}
\begin{center}
\begin{tabular}{|c|c|c|c|c|c|c|c|c|c|c|c|}
\hline        
 Scenario & $k_0$ & $k_0'$ & $d_0$ & $d_0'$ & $d_1'$ & $d_3'$ & $b'$ & $c_1$ &
 $c_2$ & $d_2$ & $d_3$  \\
\hline
 I & 1 & 0 & -1 & X & -2 & 1 & -$\frac{1}{71}$ & -$\frac{60}{71}$
 & $\frac{33}{71}$ & -$\frac{3}{71}$ & -$\frac{102}{71}$ \\
 II & 0 & 1 & X & -2 & 1 & 1 & $\frac{13}{58}$ & -$\frac{15}{29}$ &
 -$\frac{12}{29}$ & -$\frac{48}{29}$ & -$\frac{63}{116}$  \\
 III & 2 & 2 & 1 & 1 & 1 & 2 & -$\frac{16}{2995}$ & -$\frac{192}{599}$ &
 $\frac{528}{2995}$ & $\frac{384}{2995}$ &  $\frac{443}{599}$ \\
\hline
\end{tabular}
\end{center}
\end{table}

\begin{table}[ht]
\caption{Sample: rational $U(1)'$ charges of all the particles for the solution
with $k_1=0$, $k_1'=0$, $k_2=1$, $k_2'=2$, $k_3=2$ and $k_3'=1$
 in the $SU(6)$ model on $M^4\times T^2/(Z_2)^4$.
\label{tab:suv15}}
\vspace{0.4cm}
\begin{center}
\begin{tabular}{|c|c|c|c|c|c|c|c|c|c|c|c|}
\hline        
 Scenario & $k_0$ & $k_0'$ & $d_0$ & $d_0'$ & $d_2'$ & $d_3'$ & $b'$ & $c_1$ &
 $c_2$ & $d_2$ & $d_3$  \\
\hline
 I & 2 & 0 & -1 & X & 2 & 1 & -$\frac{7}{45}$ & -$\frac{4}{3}$ &
 $\frac{11}{15}$ & -$\frac{1}{15}$ & -2 \\
 II & 3 & 2 & -1 & 1 & 1 & 1 & $\frac{5}{132}$ &
 $\frac{1}{2}$ & -$\frac{1}{11}$ & $\frac{19}{22}$ &
 $\frac{13}{44}$  \\
 III & 0 & 2 & X & 1 & 1 & 1 & $\frac{1}{45}$ & $\frac{8}{15}$ &
 $\frac{1}{15}$ & -$\frac{14}{15}$ & $\frac{2}{5}$ \\
\hline
\end{tabular}
\end{center}
\end{table}

\end{document}